\documentclass[a4paper,fleqn,usenatbib]{mnras}
\usepackage{color}
\usepackage{natbib}
\usepackage{paralist}
\usepackage{graphicx}
\usepackage{epstopdf}
\usepackage{epsfig}
\usepackage{amsmath}
\usepackage{tablefootnote}

\usepackage{graphicx}	
\usepackage{amsmath}	
\usepackage{amssymb}	
\usepackage{multicol}        
\usepackage{bm}		
\usepackage{pdflscape}	

\newcommand\teff{{T_{\rm eff}}}

\newcommand\lta{\mathrel{\hbox{\raise 0.6 ex \hbox{$<$}\kern
                   -1.8 ex\lower .5 ex\hbox{$\sim$}}}}
\newcommand\gta{\mathrel{\hbox{\raise 0.6 ex \hbox{$>$}\kern
                   -1.7 ex\lower .5 ex\hbox{$\sim$}}}}
\newcommand{\bea}{\begin{eqnarray}}
\newcommand{\eea}{\end{eqnarray}}

\newcommand\species[2]{#1 {\sc #2}}
\newcommand\iso[2]{$^{\rm #1}$#2}
\newcommand{\eps}[1]{\mbox{log~$\epsilon$(#1)}}

\def\teff{\mbox{$T_{\rm eff}$}}
\def\logg{\mbox{log~{\it g}}}
\def\vmicro{\mbox{$\xi_{\rm t}$}}
\def\kmsec{\mbox{km~s$^{\rm -1}$}}

\def\eg{\mbox{e.g.}}
\def\loggf{\mbox{log~{\it gf}}}
\def\rpro{\mbox{\textit{$r$}-process}}
\def\spro{\mbox{\textit{$s$}-process}}
\def\ncap{\mbox{\textit{$n$}-capture}}
\def\carbiso{\mbox{$^{12}$C/$^{13}$C}}

\usepackage[T1]{fontenc}
\usepackage{ae,aecompl}

\usepackage{newtxtext,newtxmath}

\title[IR abundances of the RGs in the NGC 6940]{\textit{Chemical Abundances Of Open 
Clusters From High-Resolution Infrared Spectra. I. NGC 6940}}

\author[G. B\"{o}cek Topcu  et al.]{
        G.  B\"{o}cek Topcu$^{1}$\thanks{Contact e-mail: gamzebocek@gmail.com (GBT); 
        melike.afsar@ege. edu.tr (MA); chris@verdi.as.utexas.edu (CS);
        cpilacho@indiana.edu (CAP); pavelden@uvic.ca (PAD); vandenbe@uvic.ca (DAV); 
        gmace@astro.as.utexas.edu (GNM); 
        hkim@gemini.edu (HK); dtj@astro.as.utexas.edu (DTJ)}, 
        M. Af\c{s}ar$^{1,2}$, 
        C. Sneden$^{2}$, 
        C. A. Pilachowski$^{3}$,
        P. A. Denissenkov$^{4}$
\newauthor 
        D. A. VandenBerg$^{4}$,
	E. Strickland$^{2}$,
	S. \"{O}zdemir$^{1}$,
        G. N. Mace$^{2}$, 
        H. Kim$^{5}$,
        and D. T. Jaffe$^{2}$ \\ 
$^{1}$Department of Astronomy and Space Sciences, 
                 Ege University, 35100 Bornova, \.{I}zmir, Turkey \\
$^{2}$Department of Astronomy and McDonald Observatory,
                 The University of Texas, Austin, TX 78712 \\
$^{3}$Indiana University, Department of Astronomy SW319, 727 E 3rd Street,   
                 Bloomington, IN 47405 USA \\     
$^{4}$Department of Physics and Astronomy, University of Victoria,
                 Victoria, BC, V8W 2Y2, Canada            \\              
$^{5}$Gemini Observatory, Casilla 603, La Serena, Chile}

\date{Accepted 2019 March 7}

\pubyear{2018}

\begin{document}
\label{firstpage}
\pagerange{\pageref{firstpage}--\pageref{lastpage}}
\maketitle

\begin{abstract}

We present near-infrared spectroscopic analysis of 12 red giant members of 
the Galactic open cluster NGC 6940. 
High-resolution (R$\simeq$45000) and high signal-to-noise ratio (S/N $>$ 100) 
near-infrared $H$ and $K$ band spectra were gathered with the Immersion 
Grating Infrared Spectrograph (IGRINS) on the 2.7m Smith Telescope 
at McDonald Observatory. 
We obtained abundances of H-burning (C, N, O), $\alpha$ (Mg, Si, S, Ca), 
light odd-Z (Na, Al, P, K), Fe-group (Sc, Ti, Cr, Fe, Co, Ni) and 
neutron-capture (Ce, Nd, Yb) elements. 
We report the abundances of S, P, K, Ce, and Yb in NGC 6940 for the first time.
Many OH and CN features in the $H$ band were used to obtain O and N abundances.
C abundances were measured from four different features: CO molecular lines 
in the $K$ band, high excitation \ion{C}{i} lines present in both 
near-infrared and optical, CH and C$_2$ bands in the optical region.
We have also determined \carbiso~ ratios from the R-branch band heads of 
first overtone (2$-$0) and (3$-$1) $^{12}$CO and (2$-$0) $^{13}$CO lines near 
23440 \AA\ and (3$-$1) $^{13}$CO lines at about 23730 \AA. 
We have also investigated the HF feature at 23358.3 \AA, 
finding solar fluorine abundances without ruling out a slight enhancement. 
For some elements (such as the $\alpha$ group), IGRINS data yield more internally 
self-consistent abundances.
We also revisited the CMD of NGC 6940 by determining the most probable 
cluster members using $Gaia$ DR2. 
Finally, we applied Victoria isochrones and MESA models in order to refine
our estimates of the evolutionary stages of our targets.
\end{abstract}



\section{Introduction}\label{intro}

Photometry and spectroscopy of evolved stars in open
clusters contain vital information for better understanding of
stellar evolution, internal nucleosynthesis, and envelope mixing. 
Compared to field stars, cluster members can be more easily tagged in mass, 
metallicity, age, and reddening, allowing one to use high resolution
spectroscopy to derive reliable model atmospheres and surface chemical 
compositions. 
Of particular interest are the abundances of light elements and isotopes that 
can be altered through hydrogen fusion cycles: Li, C, N, O, and \carbiso.

\begin{table*}
\begin{minipage}{180mm} 
 \caption{Basic parameters of the program stars and summary of IGRINS observations.} 
 \label{tab-basic}
 \begin{tabular}{@{}lcccccccccccc@{}}
 \hline
Star     &  RA($Gaia$)   &  Dec($Gaia$)  &   V$^{\rm a}$   &   H$^{\rm b}$   &   K$^{\rm b}$   &
 G$^{\rm c}$  &  (G$_{\rm BP}$$-$G$_{\rm RP}$)$^{\rm d}$ &  $(B - V)_{0}$ & $(V - K)_{0}$ &
 Date  & Exposure   & S/N \\
      &   &     &    &    &   &   &   &  & & (UT)            &  (s) &     \\         
 \hline
MMU 28   & 20 33 25.01 &  28 00 46.8 & 11.56 &  9.05 &  8.93 &  11.22 & 
   1.33  & 0.91 & 2.06 & 31 07 2015    &  300 &  115 \\
MMU 30 &   20 33 29.80 &  28 17 04.8 & 10.89 &  8.27 &  8.18 &  10.61 &
   1.38 & 1.05 & 2.14 & 31 07 2015    &  300 &  117 \\
MMU 60 &   20 33 59.57 &  28 03 01.6 & 11.56 &  9.10 &  8.97 &  11.24 &
  1.31 & 0.90 & 2.02 &  31 07 2015    &  300 &   94 \\
MMU 69 &   20 34 05.74 &  28 11 18.2 & 11.64 &  9.07 &  8.95 &  11.28 &
  1.34 & 0.87 & 2.12 &  01 08 2015    & 300  &  126 \\
MMU 87 &   20 34 14.70 &  28 22 15.7 & 11.32 &  8.87 &  8.74 &  11.00 &
 1.30 & 0.87 & 2.01 &   01 08 2015    & 300  &  116 \\
MMU 101&   20 34 23.64 &  28 24 25.5 & 11.27 &  8.87 &  8.74 &  10.96 &
  1.28 &  0.89 & 1.96 & 01 08 2015    & 300  &  155 \\
MMU 105&   20 34 25.46 &  28 05 05.5 & 10.66 &  7.95 &  7.82 &  10.30 &
  1.40 & 1.01 & 2.27 &  06 08 2015    & 300  &  154 \\
MMU 108&   20 34 25.68 &  28 13 41.5 & 11.19 &  8.86 &  8.70 &  10.88 &
  1.27 & 0.83 & 1.92 &  01 08 2015    & 300  &  113 \\
MMU 132&   20 34 40.11 &  28 26 38.8 & 10.97 &  8.49 &  8.39 &  10.65 &
  1.29 & 0.89 & 2.01 &  08 08 2015    & 300  &  139 \\
MMU 138&   20 34 45.87 &  28 09 04.6 & 11.36 &  8.96 &  8.81 &  11.05 &
  1.28  &  0.87 & 1.98 &   16 06 2016    & 420  &  129 \\
MMU 139&   20 34 47.60 &  28 14 47.1 & 11.38 &  8.94 &  8.80 &  11.05 &
 1.28 &  0.89 & 2.02 & 08 08 2015    & 300  &  156 \\
MMU 152&   20 34 56.64 &  28 14 27.0 & 10.84 &  8.35 &  8.26 &  10.50 &
 1.27 &  0.87 & 2.01 & 16 06 2016    & 300  &  105 \\
 \hline
\multicolumn{9}{l}{$^{\rm a}$ \cite{lar60}} \\
\multicolumn{9}{l}{$^{\rm b}$ \cite{2MASS}}\\
\multicolumn{9}{l}{$^{\rm c}$ \cite{GAIA18b}}\\
\multicolumn{9}{l}{$^{\rm d}$ \cite{cantat18}}\\
\end{tabular}
\end{minipage}
\end{table*}

Open star cluster formation theories generally 
assume chemically homogeneous molecular clouds.
With few exceptions (e.g. for M67 see \citealt{souto18,bertelli18,gao18,randich06}), 
detailed abundances for main sequence (MS) stars have not 
been derived for the kinds of intermediate/old open clusters that have 
well-developed red giant branches.
Therefore the default assumption is that for solar metallicity\footnote{
We adopt the standard spectroscopic notation 
\citep{wallerstein59} that for elements A and B,
[A/B] $\equiv$ log$_{\rm 10}$(N$_{\rm A}$/N$_{\rm B}$)$_{\star}$ $-$
log$_{\rm 10}$(N$_{\rm A}$/N$_{\rm B}$)$_{\odot}$.
We use the definition
\eps{A} $\equiv$ log$_{\rm 10}$(N$_{\rm A}$/N$_{\rm H}$) + 12.0, and
equate metallicity with the stellar [Fe/H] value.}
clusters (which encompass the vast majority of known cases), the natal 
light element abundance ratios are equal to the solar ones.
Departures from the solar ratios in evolved cluster members are assumed to 
be due to stellar evolutionary effects.

LiCNO abundances of open cluster red giants usually 
have been derived from features present in high-resolution 
optical\footnote{
In this paper we will use ``optical'' or ``opt'' to 
refer to the spectral domain $\lambda$~$<$~9000~\AA, and ``$IR$'' to refer 
to the $H$ (1.5-1.8~$\micron$) and $K$ (2.0-2.4~$\micron$) near-$IR$ 
photometric bands.  
For internal consistency we will use \AA\ wavelength units throughout.}
wavelength spectra: C from CH and C$_2$ bands, N from 
CN bands, O from the [\ion{O}{i}] 6300~\AA\ line, and Li from the 
\ion{Li}{i} 6707~\AA\ resonance doublet.
All of these spectral features have assets and liabilities.
For example, the single O abundance indicator\footnote{
Other O abundance features all have observational and analytic difficulties.  
The [\ion{O}{i}] 6363.8~\AA\ line is blended with CN and lies in a 
general flux depression caused by the \ion{Ca}{i} 6361.8~\AA\ 
auto ionization line. 
The \ion{O}{i} 7770~\AA\ triplet is subject to
large non-local thermodynamic equilibrium (NLTE) corrections.
The OH electronic transition bands occur at wavelengths below 3300~\AA,
a low-flux complex line region that is very difficult to be studied from the ground.}
has a significant \ion{Ni}{i} blend and lies near telluric
O$_2$ and night sky emission lines.  
The CH G band is often very strong and 
it occurs in
the heavily contaminated 4300~\AA\
region, while the C$_2$ Swan bands near 4720, 5170, and 5630~\AA\ are weak
and blended.
CN transitions can easily be observed throughout the red region, but are 
mostly very weak except for $\lambda$~$>$~8000~\AA; N abundances from CN
also are completely dependent on derived C abundances.

The $IR$ wavelength domain contains several species 
that can greatly improve on CNO abundances determined from optical spectra.
OH vibration-rotation band lines occur throughout the 1$-$2~$\mu$m 
region, as do stronger lines of the CN red electronic system.  
There are many high-excitation \ion{C}{i} transitions in the $IR$
spectral region.
Analyses of these CNO abundance indicators together with those in the
optical region should yield reliable abundances for use in stellar evolution
studies.

High-resolution spectroscopy of globular cluster
red giants in the $H$ band has been conducted 
over the last couple of decades with Keck/NIRSPEC \citep{mclean98}
and VLT/CRIRES \citep{kaeufl04} instruments (e.g. 
\citealt{origlia02,origlia04,origlia05,origlia08,valenti11,delaverny13,valenti15}). 
The detailed chemical abundances of some Galactic globular clusters were also
studied by \cite{lamb15} using high-resolution spectral data obtained both in the
optical and $H$ band regions. Most recently, \cite{lamb17} reported the $IR$ abundances
for metal-poor stars in and towards the Galactic Centre from $H$ band spectra taken with
the RAVEN multi-object science demonstrator and IRCS/Subaru instrument.
A few chemical composition studies of open cluster members have been 
performed with the Keck/NIRSPEC and VLT/CRIRES (e.g. \citealt{origlia06,maiorca14}) 
as well as with TNG/Giano (\citealt{oliva12} and references therein); see
for example \cite{origlia13,origlia16}.
Elemental abundances for a limited number of open clusters have been 
investigated using the $H$ band spectral data of APOGEE 
(R~$\simeq$~25,000, \citealt{majewski17}); 
see for example
\cite{cunha15}, \cite{souto16}, \cite{linden17}, \cite{souto18}.

In this paper we present an abundance study of the
intermediate age open cluster (OC) NGC~6940, using high resolution 
$IR$ spectra.  
This work is part of a series of studies that involve chemical composition
analyses of selected red giant (RG) members of OCs from high-resolution 
spectra obtained in both optical and $IR$ spectral regions.
The results from the optical spectral range of the 12 RG members of NGC~6940
have been previously reported by \cite{bocek16} (hereafter Paper~1). 
In this study, we present abundances obtained from high-resolution 
$H$ and $K$ band spectra for the same RG members of NGC~6940. 
To the best of our knowledge, this study is the first of its kind that 
reports internally-consistent abundances for a large number of OC red giants
using high resolution spectra in the optical (5100$-$8800 \AA) and
the complete $H$ and $K$ bands.
Our investigation puts special emphasis on determining better CNO abundances
towards interpretation of H-fusion synthesis and mixing in open clusters.

\section{Observations and Data Reduction}\label{obs}

High-resolution 
($R\equiv\lambda/\Delta\lambda\approx45.000$)
and high S/N spectra of the NGC~6940 targets 
were obtained with the Immersion Grating Infrared Spectrometer 
(IGRINS, \citealt{yuk10,park14,mace16}) on the 2.7m Harlan J. Smith Telescope 
at McDonald Observatory. 
Observations were made between July 2015 and June 2016. 
Stellar coordinates, broad-band photometric data, $Gaia$ magnitudes, 
colours, and the observation log are given in Table~\ref{tab-basic}.

IGRINS has total wavelength coverage of 1.48$-$2.48 $\mu$m with
only about 100~\AA\ lost in the middle, unobservable 
from the ground due to heavy telluric absorption.
This large wavelength coverage is obtained in a single observation per star,
with the $H$ and $K$ spectral regions being captured by different 
camera/detector systems.\footnote{
See Figure~1 at
\url{http://www.as.utexas.edu/astronomy/research/people/jaffe/igrins.html}}
Flat field, ThAr lamp and sky calibration frames were taken during each night.
The spectra of targets and of hot, rapidly-rotating telluric standards 
were taken by nodding the stars along the slit with a four-integration 
ABBA pattern.

To reduce the data frames we used the IGRINS reduction 
pipeline package PLP2 \citep{lee15}.\footnote{\url{
https://github.com/igrins/plp}}
The PLP2 software applies basic echelle reduction steps such as sky 
subtraction, flat fielding, bad pixel correction, background removal, 
aperture extraction, and wavelength calibration. 
Wavelength solutions were derived from the Th-Ar lamp lines and then 
improved by using sky OH emission lines. 
To remove the contamination of CO$_{2}$ lines we have used the
\textit{telluric} task of IRAF\footnote{\url{http://iraf.noao.edu/}}.
More detail about the reduction process can be found in \cite{afsar16}.

\begin{table*}                                                                                                   
 \begin{minipage}{135mm}                                      
 \caption{Kinematics and radial velocities.}
 \label{tab-motions}                                            
 \begin{tabular}{@{}lcccccc@{}}                             
 \hline                                                       
Star     &  $\pi_{\rm (\it Gaia)}$  & $\mu_\alpha$ $_{\rm (\it Gaia)}$ & $\mu_\delta$ $_{\rm 
(\it Gaia)}$ & RV$^{\rm a}$  & RV$_{\rm (Paper~1)}$   &   RV$_{\rm (\it Gaia)}$  \\
   & (mas yr$^{-1}$) &  (mas yr$^{-1}$)   &  (mas yr$^{-1}$)   &
  (km s$^{-1}$) &  (km s$^{-1}$) & (km s$^{-1}$)               \\
 \hline
MMU 28   & 1.001$\pm$0.033 & $-$2.015$\pm$0.047 & $-$9.523$\pm$0.055 & 
   8.24$\pm$0.25 &
   8.90$\pm$0.22 & 8.70$\pm$0.37  \\
MMU 30   & 0.958$\pm$0.028 & $-$1.919$\pm$0.042 & $-$9.382$\pm$0.048 &
   7.46$\pm$0.15                         &
   7.96$\pm$0.20 & 8.03$\pm$0.25  \\
MMU 60   & 0.977$\pm$0.029 & $-$2.240$\pm$0.042 & $-$9.263$\pm$0.041 & 
   7.32$\pm$0.25                          &
   7.66$\pm$0.22 & 7.41$\pm$0.32  \\
MMU 69   & 0.930$\pm$0.027 & $-$2.046$\pm$0.042 & $-$9.329$\pm$0.038 &
   7.54$\pm$0.25                          &
   8.08$\pm$0.24 & 8.21$\pm$0.34  \\
MMU 87   & 0.998$\pm$0.032 & $-$1.985$\pm$0.044	& $-$9.462$\pm$0.040 &
   8.06$\pm$0.26                          &
   7.98$\pm$0.27 & 8.13$\pm$0.30  \\
MMU 101  & 1.000$\pm$0.035 & $-$1.973$\pm$0.051 & $-$9.460$\pm$0.046 &
   7.36$\pm$0.21                          &
   7.74$\pm$0.23 & 6.92$\pm$0.38  \\
MMU 105  & 0.979$\pm$0.031 & $-$1.955$\pm$0.042 & $-$9.346$\pm$0.043 &
   7.61$\pm$0.29                          &
   7.74$\pm$0.23 & 8.38$\pm$0.23  \\
MMU 108  & 0.883$\pm$0.033 & $-$1.534$\pm$0.045 & $-$9.235$\pm$0.041 &
   7.51$\pm$0.34                         &
   7.39$\pm$0.25 & 8.10$\pm$0.50  \\
MMU 132  & 1.005$\pm$0.038 & $-$2.015$\pm$0.051 & $-$9.314$\pm$0.047 &
   7.48$\pm$0.25                         &
   7.76$\pm$0.42 & 7.62$\pm$0.28  \\
MMU 138  & 0.920$\pm$0.029 & $-$1.803$\pm$0.042 & $-$9.457$\pm$0.040 &
   8.12$\pm$0.32                          &
   8.22$\pm$0.23 & 7.90$\pm$0.30  \\
MMU 139  & 0.975$\pm$0.031 & $-$2.046$\pm$0.045 & $-$9.523$\pm$0.046 &
   7.97$\pm$0.21                         &
   7.53$\pm$0.23 & 8.25$\pm$0.32  \\
MMU 152  & 0.982$\pm$0.032 & $-$1.924$\pm$0.044 & $-$9.378$\pm$0.041 &
   8.91$\pm$0.24                         &
   9.28$\pm$0.24 & 9.32$\pm$0.21 \\
         &       &         &     &           &            &          \\
mean     & 0.967$\pm$0.011 & $-$1.955$\pm$0.048 & $-$9.389$\pm$0.028 &  
   7.80$\pm$0.14 & 7.91$\pm$0.12 & 8.08$\pm$0.18                   \\
$\sigma$ &           0.038 &              0.168 &              0.096 &  
     0.47   &    0.41 &          0.61    \\
 \hline
\multicolumn{5}{l}{$^{\rm a}$ This study.} \\
\end{tabular}
\end{minipage}                                                
\end{table*}

\section{Kinematics, Distances, and Luminosities}\label{kinematics}

\begin{figure}
\includegraphics[width=\columnwidth]{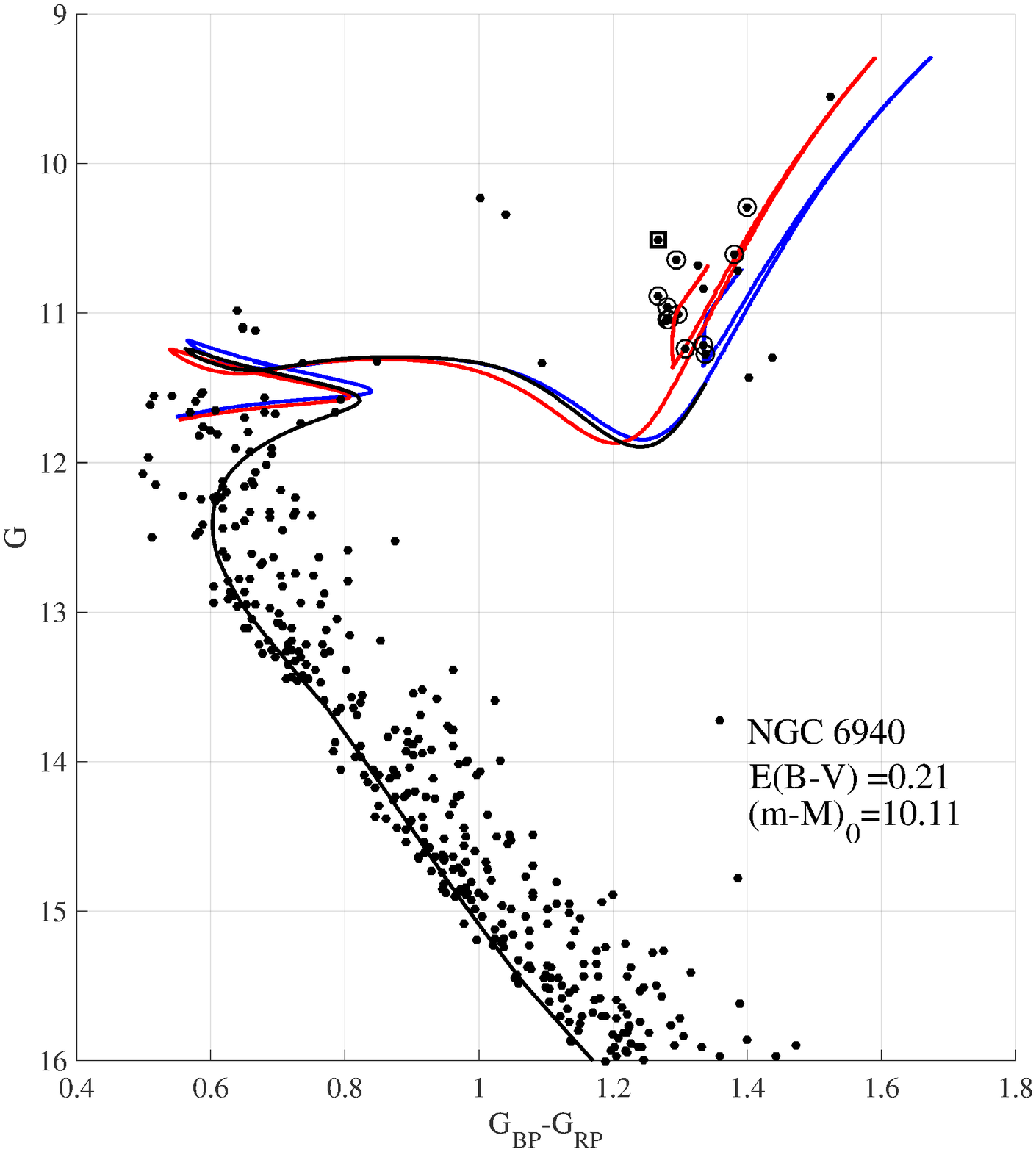}
\caption{The CMD of NGC\,6940 (black dots) and its fit 
         with a new $1.15$ Gyr isochrone (black curve) generated with the 
         Victoria stellar evolutionary code \citep{vandenberg:14}. 
         The blue and red 
         curves are the MESA \citep{paxton:11,paxton:13} evolutionary tracks 
         both computed for M = 2 M$_\odot$, [Fe/H] $ = 0$, $Y = 0.27$
         but different assumptions about the mixing length parameter.
         See \S\ref{isoch} for a detailed descriptions of these theoretical
         curves. 
         Circled symbols represent our targets. 
         The location of MMU 152, a star with an unusual light element abundance set,
         is framed with a square.
        }
\label{NGC6940GaiaDR2}
\end{figure}

The second data release (DR2) of the $Gaia$ astrometric 
satellite mission \citep{GAIA16,GAIA18b} includes kinematic information 
for all of our NGC~6940 RG stars.
In Table~\ref{tab-motions} we list $Gaia$ parallaxes, proper motions,
and radial velocities (RVs) 
along with the RVs measured from both $IR$ and optical 
(Paper~1) spectra.
$IR$ RVs were measured in the same manner as described in Paper~1
using the regions that are less affected by the atmospheric lines.
All radial velocities are in excellent agreement within about 0.5~\kmsec. 
The RVs from our data set yield a cluster mean of
$\langle$RV$_{\rm mean}$$\rangle$~$\simeq$~7.91~$\pm$~0.14 ($\sigma$~=~0.49)~\kmsec, 
and a star-by-star comparison with $Gaia$ yields 
$\langle$RV$_{\rm \it Gaia}$ $-$ RV$_{\rm mean}$$\rangle$~= 0.17~$\pm$~0.11 ($\sigma$~=~0.39).
In all data sets MMU 152 stands out from the cluster mean: 
$\langle$RV$_{\rm MMU152}$$\rangle$~$\simeq$~9.20~\kmsec, slightly outside the 
observational uncertainties.
However, its parallax and proper motions are well within the cluster means, 
confirming its membership in NGC 6940.

Paper~1 investigated the color magnitude diagram (CMD) of NGC 6940
by using the data provided by in WEBDA\footnote{\url{
http://www.univie.ac.at/webda/webda.html}}
database (\citealt{walker58,lar60,hoag61,stetson00}). 
Here we revisit the CMD and gather the data from \cite{cantat18} who studied 
the open cluster population in the Milky Way by making use of $Gaia$ DR2. 
Among the members listed in \citeauthor{cantat18}, we only used the ones 
with membership probabilities $\geq$ 80\%. We plot the $Gaia$ CMD in 
Figure~\ref{NGC6940GaiaDR2}. We illustrated our targets by 
framing their locations with circles and one square for MMU 152.
The mean parallax of the cluster from these stars is
$\langle$$\pi$$_{\rm (\it Gaia)}$$\rangle$~= 0.949~$\pm$~0.002 ($\sigma$~=~0.052)
$mas$ (miliarcsec), which is about 0.350~$mas$ less than the parallax we adopted 
from \cite{khar05} in Paper 1.
This difference cannot be due to systematics in $Gaia$ parallaxes,
which can be up to 0.03 $mas$ throughout the sky as recently investigated by \cite{aren18}
in detail. 
Figure~\ref{NGC6940GaiaDR2} also contains theoretical isochrones and evolutionary 
tracks that are needed for interpretation of the chemical compositions
derived in this study. The true distance modulus (m-M)$_{\rm 0}$ = 10.11
information for NGC 6940 is also given in the Figure along with 
$E(B-V) = 0.21$, which is the reddening estimate obtained from the fit of 
isochrones to the MS stars > 1.5 mag below the cluster turnoff when the
distance modulus based on $Gaia$ parallaxes is adopted.
These curves need extended discussion, which we defer to \S\ref{isoch}.

\begin{table*}
 \begin{minipage}{173mm}
 \caption{Model atmosphere parameters.} 
 \label{tab-model}
  \begin{tabular}{@{}lcccccrccrccrcc@{}}
  \hline
 Star     &  \teff\ & \teff$_{(\rm \it Gaia)}$ &  \teff$_{\rm (LDR)}$ & \logg &     $\xi_{t}$ & 
     [\ion{Fe}{i}/H]  & $\sigma$ & \# & 
     [\ion{Fe}{ii}/H] & $\sigma$ & \# &
     [\ion{Fe}{i}/H]  & $\sigma$ & \# \\
          &    (K)  &        (K)  &    (K) &   & (km s$^{-1}$) &
                    (opt) &    (opt) &   (opt) &
                    (opt) &    (opt) &   (opt) &
                     ($IR$) &     ($IR$) &    ($IR$) \\
 \hline
MMU 28    &   5024 &  4742 & 5052 & 2.89 &  1.03 & $-$0.03 &  0.07 &    56 & $-$0.07 &  0.05 &    11 &    0.00 &  0.04 &    19 \\ 
MMU 30    &   4959 &  4706 &  4977 & 2.85 &  1.32 &    0.01 &  0.07 &    53 & $-$0.04 &  0.08 &    10 &    0.00 &  0.04 &    24 \\  
MMU 60    &   5046 &  4683 &  5048 & 2.97 &  0.97 &    0.04 &  0.06 &    50 &    0.01 &  0.08 &     9 &    0.06 &  0.05 &    20 \\ 
MMU 69    &   5004 &  4649 &  5068 & 2.90 &  1.05 & $-$0.03 &  0.06 &    55 & $-$0.07 &  0.08 &    10 &    0.03 &  0.05 &    24 \\
MMU 87    &   5023 &  4692 &  5048 & 2.85 &  1.07 &    0.05 &  0.07 &    50 &    0.03 &  0.06 &    10 &    0.07 &  0.04 &    25 \\ 
MMU 101   &   5037 &  4767 &  5035 & 3.02 &  1.16 &    0.01 &  0.07 &    55 & $-$0.02 &  0.06 &     8 &    0.05 &  0.04 &    23 \\
MMU 105   &   4765 &  4580 &  4893 & 2.34 &  1.35 & $-$0.10 &  0.07 &    53 & $-$0.15 &  0.08 &     8 & $-$0.04 &  0.06 & 24 \\
MMU 108   &   5132 &  4802 &  5082 & 2.80 &  1.28 & $-$0.15 &  0.07 &    54 & $-$0.22 &  0.03 &     9 & $-$0.06 &  0.06 & 18 \\
MMU 132   &   4962 &  4822 &  4972 & 2.65 &  1.29 & $-$0.01 &  0.05 &    51 & $-$0.09 &  0.09 &    10 &    0.09 &  0.05 &   24 \\
MMU 138   &   5056 &  4703 &  5073 & 3.00 &  1.10 &    0.00 &  0.06 &    47 & $-$0.05 &  0.04 &     8 &    0.03 &  0.08 &    21 \\         
MMU 139   &   5013 &  4789 &  4958 & 2.99 &  1.10 &    0.01 &  0.07 &    51 & $-$0.03 &  0.05 &     8 &    0.06 &  0.06 &    23 \\
MMU 152   &   4933 &  4846 & 4941 & 2.66 &  1.36 & $-$0.06 &  0.06 &    52 & $-$0.13 &  0.06 &    10 &    0.00 &  0.07 &    19 \\
 \hline
\end{tabular}
\end{minipage}
\end{table*}

\section{Model Atmospheres and Abundances from the Optical Window}\label{modopt}

In Paper~1 we reported model atmospheric parameters \teff, \logg, \vmicro, 
[Fe/H], absolute abundances \eps{Li}, the \carbiso\ ratios, and relative 
abundances [X/Fe] of 22 elements for 12 NGC~6940 RG members 
(see Table~9 in Paper~1).
Here we additionally report the abundances of S and Ce from their neutral
and ionized species, respectively, and determine C abundances also from high excitation 
\ion{C}{i} lines (Table~\ref{tab-abunds}). 
We also reanalyse the \ion{Sc}{ii} lines adopting the \loggf\
values from \cite{lawler89}.
The abundance analyses were carried out using high-resolution, high-S/N 
optical spectra obtained with the 9.2-m Hobby-Eberly Telescope (HET) 
High-Resolution 
Spectrograph (HRS; \citealt{tull98}) at the McDonald Observatory. 
We used the local thermodynamic equilibrium (LTE) line analysis and 
synthetic spectrum code MOOG \citep{sneden73}\footnote{\url{
http://www.as.utexas.edu/~chris/moog.html} }
to derive the model atmospheric parameters (\teff, \logg, [Fe/H] and 
$\xi_{t}$) and chemical abundances.
The model atmospheric parameters were determined from equivalent width ($EW$
{\footnote{Equivalent width and line depth ratio (LDR) 
measurements were completed in the same manner described in Paper 1.}}) 
measurements of \ion{Fe}{i}, \ion{Fe}{ii}, \ion{Ti}{i},
and \ion{Ti}{ii} lines (Table~6 of Paper~1).
Those parameters are adopted without change in this study and are listed 
in Table~\ref{tab-model}.

An extended description of the optical abundance 
analyses is given in Paper~1, in which we derived solar photospheric 
abundances in the same manner as those in the program stars, making in 
effect the stellar abundances into differential values relative to the Sun.
In the present study we have adopted the solar abundances of 
\cite{asplund09}. 
While most of the optical region lines have reliable laboratory transition 
probabilities, many more lines in our $IR$ spectra must have 
transition probabilities derived from reverse solar analysis.
Therefore we adopt uniformly the \citeauthor{asplund09} solar values 
in this work.
In the upper part of Table~\ref{tab-abunds} we list the relative optical
abundance ratios using this assumed set of solar abundances.
In this paper we will examine the LiCNO group abundances 
in more detail than was done in Paper~1, including a re-assessment of the 
values derived from optical data.
In Table~\ref{tab-abunds} these abundances and \carbiso\ ratios are entered 
at the ends of the rows for optical and IR results, and will be discussed 
separately in \S\ref{cnoiso}. 
The [X/Fe] calculations were performed by taking
species by species differences using both \ion{Fe}{i} and  
\ion{Fe}{ii} abundances.

\section{Temperature determination from the $IR$ region}\label{ldrcomp}

\begin{figure}
 \includegraphics[width=\columnwidth]{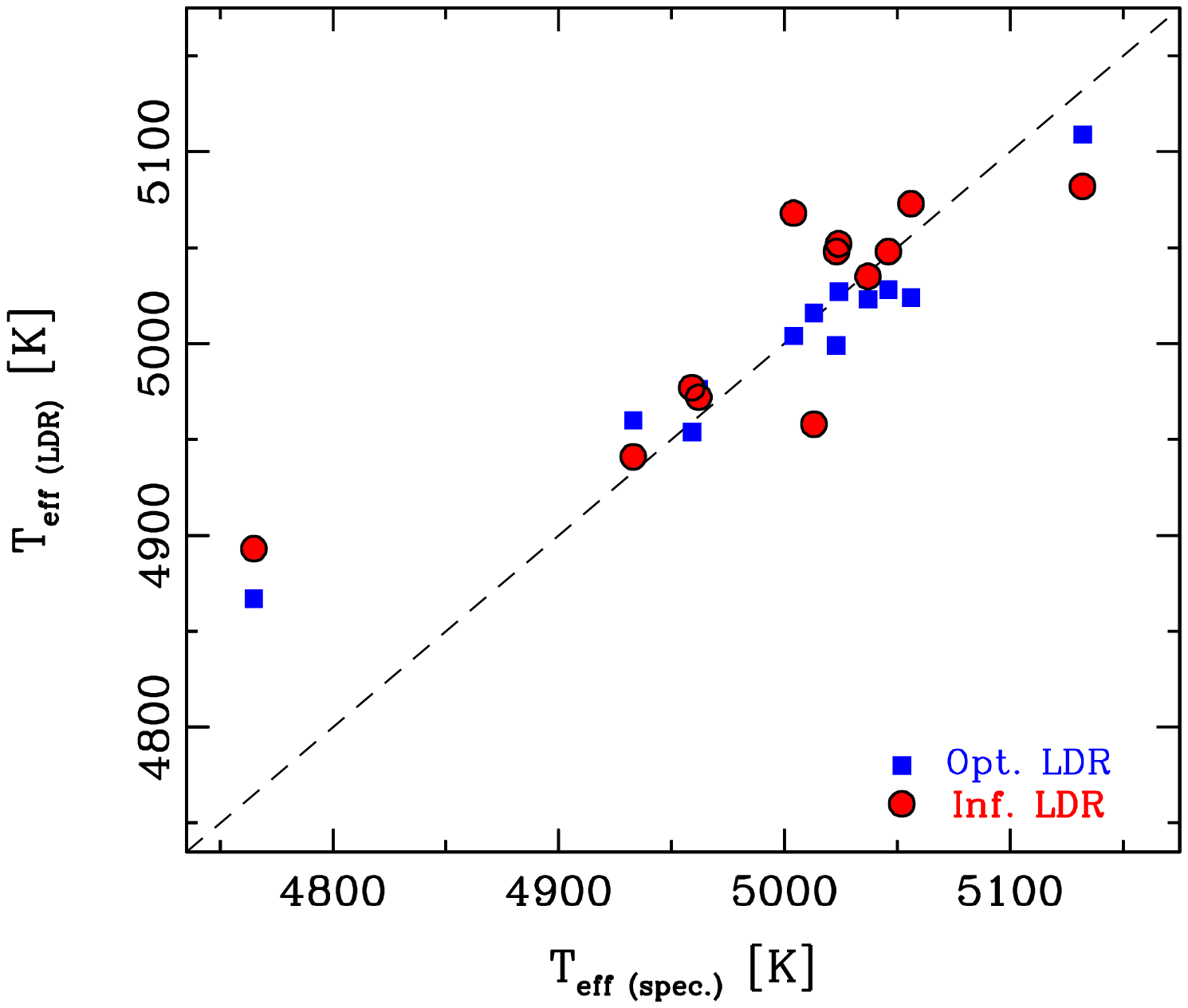}
      \caption{Correlation of optical and $H$-band LDR \teff\ values 
               with those derived spectroscopically in Paper~I.
               The dashed line represents equality of the temperatures.}
      \label{teff}
\end{figure}

\begin{table*}
  \caption{Relative abundances and \carbiso\ ratios of NGC 6940 RGs and their cluster means.}
   \label{tab-abunds}
  \begin{tabular}{@{}lrrrrrrrrrrrrrcr@{}}
  \hline
Species  & \multicolumn{12}{c}{MMU} \\
$\rm{[X/Fe]}$& 28  &      30  &      60  &      69  &      87  &     101  &     105  &     108  &     132  &     138   &     139 &      152 &     mean & $\sigma$ & \#$_{max}$ \\ 
\hline
\multicolumn{15}{c}{Optical Spectral Region} \\
\ion{Na}{i}	&	0.23	&	0.22	&	0.19	&	0.24	&	0.12	&	0.24	&	0.34	&	0.45	&	0.34	&	0.16	&	0.20	&	0.53	&	0.27	&	0.12	&	4	\\
\ion{Mg}{i}	&	0.06	&	0.01	&	0.05	&	0.05	&	0.03	&	0.08	&	0.10	&	0.17	&	0.09	&	0.07	&	0.05	&	0.13	&	0.07	&	0.04	&	2	\\
\ion{Al}{i}	&$-$0.07	&$-$0.10	&$-$0.12	&$-$0.03	&$-$0.07	&$-$0.08	&$-$0.02	&	0.09	&$-$0.02	&$-$0.06	&$-$0.05	&	0.05	&$-$0.04	&	0.06	&	2	\\
\ion{Si}{i}	&	0.21	&	0.23	&	0.12	&	0.22	&	0.19	&	0.24	&	0.27	&	0.22	&	0.29	&	0.16	&	0.24	&	0.30	&	0.22	&	0.05	&	15	\\
\ion{S}{i}$^{*}$	&	0.13	&	0.11	&	0.04	&	0.12	&	0.08	&	0.10	&	0.15	&	0.21	&	0.19	&	0.15	&	0.15	&	0.25	&	0.16	&	0.06	&	2	\\
\ion{Ca}{i}	&	0.13	&	0.07	&	0.12	&	0.16	&	0.10	&	0.09	&	0.15	&	0.18	&	0.14	&	0.05	&	0.12	&	0.11	&	0.14	&	0.04	&	10	\\
\ion{Sc}{ii}$^{*}$	&	0.11	&	0.16	&	0.14	&	0.24	&	0.11	&	0.23	&	0.16	&	0.13	&	0.13	&	0.16	&	0.17	&	0.24	&	0.16	&	0.05	&	6	\\
\ion{Ti}{i}	&$-$0.06	&$-$0.09	&$-$0.06	&$-$0.08	&$-$0.10	&$-$0.07	&$-$0.08	&	0.00	&$-$0.05	&$-$0.01	&$-$0.09	&$-$0.04	&$-$0.06	&	0.03	&	11	\\
\ion{Ti}{ii}	&	0.07	&	0.14	&	0.07	&	0.08	&	0.06	&	0.10	&	0.09	&	0.15	&	0.16	&	0.11	&	0.18	&	0.09	&	0.11	&	0.04	&	4	\\
\ion{V}{i}	&$-$0.03	&$-$0.03	&$-$0.07	&$-$0.05	&$-$0.10	&$-$0.02	&$-$0.06	&	0.03	&$-$0.03	&$-$0.01	&$-$0.07	&$-$0.01	&$-$0.04	&	0.04	&	12	\\
\ion{Cr}{i}	&	0.06	&	0.04	&	0.06	&	0.10	&	0.04	&	0.05	&	0.07	&	0.12	&	0.06	&	0.10	&	0.01	&	0.06	&	0.06	&	0.03	&	14	\\
\ion{Cr}{ii}	&	0.23	&	0.29	&	0.19	&	0.17	&	0.22	&	0.21	&	0.27	&	0.20	&	0.39	&	0.17	&	0.26	&	0.23	&	0.24	&	0.06	&	3	\\
\ion{Mn}{i}	&$-$0.06	&$-$0.08	&$-$0.07	&$-$0.09	&$-$0.11	&$-$0.08	&$-$0.15	&$-$0.06	&$-$0.03	&$-$0.07	&$-$0.10	&$-$0.03	&$-$0.08	&	0.03	&	3	\\
\ion{Co}{i}	&$-$0.12	&$-$0.09	&$-$0.15	&$-$0.13	&$-$0.17	&$-$0.08	&$-$0.14	&$-$0.11	&$-$0.11	&$-$0.11	&$-$0.12	&$-$0.07	&$-$0.12	&	0.03	&	5	\\
\ion{Ni}{i}	&	0.09	&	0.09	&	0.07	&	0.09	&	0.08	&	0.11	&	0.11	&	0.06	&	0.08	&	0.08	&	0.14	&	0.10	&	0.09	&	0.02	&	29	\\
\ion{Cu}{i}	&	0.01	&	0.07	&$-$0.03	&$-$0.02	&$-$0.04	&	0.01	&$-$0.06	&$-$0.02	&	0.04	&	0.02	&	0.04	&	0.11	&	0.01	&	0.05	&	1	\\
\ion{Zn}{i}	&	0.08	&	0.08	&	0.12	&	0.10	&	0.03	&	0.10	&	0.18	&	0.04	&	0.09	&	0.05	&	0.20	&	0.03	&	0.09	&	0.05	&	1	\\
\ion{Y}{ii}	&	0.12	&	0.15	&	0.07	&	0.12	&	0.06	&	0.10	&	0.08	&	0.22	&	0.07	&	0.09	&	0.16	&	0.14	&	0.12	&	0.05	&	4	\\
\ion{La}{ii}	&	0.20	&	0.26	&	0.12	&	0.24	&	0.18	&	0.29	&	0.16	&	0.23	&	0.21	&	0.25	&	0.21	&	0.22	&	0.21	&	0.05	&	4	\\
\ion{Ce}{ii}$^{*}$	&	0.12	&	0.18	&	0.08	&	0.13	&	0.07	&	0.13	&	0.03	&	0.09	&	0.06	&	0.16	&	0.15	&	0.17	&	0.11	&	0.05	&	4	\\
\ion{Nd}{ii}	&	0.26	&	0.32	&	0.22	&	0.23	&	0.29	&	0.32	&	0.23	&	0.21	&	0.25	&	0.23	&	0.25	&	0.32	&	0.26	&	0.04	&	3	\\
\ion{Eu}{ii}	&	0.07	&	0.16	&	0.11	&	0.14	&	0.08	&	0.15	&	0.14	&	0.15	&	0.13	&	0.20	&	0.15	&	0.15	&	0.13	&	0.04	&	2	\\
log~$\epsilon$(Li) &
              1.05 &   $<$0.0 &     1.23 &     1.29 &     0.56 &     0.64 &   $<$0.1 &   $<$0.0 &     0.33 &     1.05 &     0.86 &   $<$0.0 &          &          &  1 \\
\carbiso&       25 &       15 &       20 &       10 &       25 &       15 &       15 &       20 &       25 &       20 &       15 &        6 &       18 &        6 & CN   \\
C	&$-$0.19	&$-$0.20	&$-$0.30	&$-$0.24	&$-$0.22	&$-$0.24	&$-$0.27	&$-$0.17	&$-$0.27	&$-$0.15	&$-$0.23	&$-$0.41	&$-$0.24	&	0.07	&	C$_2$, \ion{C}{i}\\
N	&	0.42	&	0.41	&	0.47	&	0.48	&	0.42	&	0.48	&	0.41	&	0.58	&	0.42	&	0.47	&	0.39	&	0.55	&	0.46	&	0.06	&	CN	\\
O	&$-$0.08	&$-$0.05	&$-$0.17	&$-$0.07	&$-$0.04	&$-$0.05	&$-$0.14	&$-$0.09	&$-$0.22	&	0.00	&$-$0.05	&$-$0.08	&$-$0.09	&	0.06	&	[\ion{O}{i}]	\\
        &          &          &          &          &          &          &          &          &          &          &          &          &          &          &    \\
\multicolumn{15}{c}{IGRINS H \& K  Spectral Region} \\
\ion{Na}{i}    &     0.28 &     0.29 &     0.16 &     0.27 &     0.20 &     0.23 &     0.34 &     0.51 &     0.18 &     0.25 &     0.15 &     0.54 &     0.28 &     0.13 &  5 \\
\ion{Mg}{i}    &     0.05 &     0.01 &     0.00 &     0.05 &     0.00 &  $-$0.06 &     0.02 &     0.07 &  $-$0.02 &     0.02 &  $-$0.02 &     0.07 &     0.02 &     0.04 & 11 \\
\ion{Al}{i}    &     0.11 &     0.09 &     0.05 &     0.05 &     0.06 &     0.08 &     0.14 &     0.06 &  $-$0.07 &  $-$0.02 &     0.03 &     0.21 &     0.07 &     0.07 &  6 \\
\ion{Si}{i}    &     0.13 &     0.16 &     0.13 &     0.14 &     0.14 &     0.13 &     0.16 &     0.17 &     0.12 &     0.16 &     0.14 &     0.19 &     0.15 &     0.02 & 11 \\
\ion{P}{i}     &     0.05 &  $-$0.05 &     0.02 &     0.09 &  $-$0.08 &     0.16 &     0.15 &     0.21 &     0.25 &  $-$0.05 &     0.13 &     0.29 &     0.10 &     0.12 &  2 \\
\ion{S}{i}     &     0.08 &     0.11 &     0.07 &     0.08 &  $-$0.02 &     0.06 &     0.08 &     0.11 &     0.06 &     0.06 &     0.03 &     0.11 &     0.07 &     0.04 & 10 \\
\ion{K}{i}     &     0.00 &     0.03 &     0.21 &     0.13 &     0.07 &     0.13 &     0.22 &     0.45 &     0.10 &     0.17 &     0.15 &     0.39 &     0.17 &     0.13 &  2 \\
\ion{Ca}{i}    &     0.21 &     0.15 &     0.08 &     0.16 &     0.15 &     0.09 &     0.19 &     0.16 &     0.13 &     0.14 &     0.14 &     0.16 &     0.15 &     0.04 & 11 \\
\ion{Sc}{i}   &     0.06 &  $-$0.04 &     0.17 &     0.16 &  $-$0.07 &  $-$0.10 &  $-$0.18 &     0.09 &     0.00 &  $-$0.08 &  $-$0.08 &     0.05 &     0.00 &     0.11 &  2 \\
\ion{Ti}{i}    &  $-$0.03 &     0.04 &  $-$0.02 &     0.02 &     0.00 &     0.01 &  $-$0.03 &     0.08 &  $-$0.02 &     0.10 &     0.00 &     0.03 &     0.01 &     0.04 & 10 \\
\ion{Ti}{ii}   &  $-$0.02 &     0.01 &  $-$0.08 &  $-$0.15 &  $-$0.08 &  $-$0.11 &  $-$0.06 &  $-$0.03 &  $-$0.07 &     0.05 &  $-$0.02 &  $-$0.04 &  $-$0.05 &     0.05 &  1 \\
\ion{Cr}{i}    &     0.07 &     0.03 &     0.01 &  $-$0.03 &  $-$0.03 &  $-$0.04 &     0.07 &     0.03 &  $-$0.03 &  $-$0.02 &     0.01 &     0.09 &     0.01 &     0.05 &  3 \\
\ion{Co}{i}    &     0.00 &     0.01 &  $-$0.13 &  $-$0.01 &  $-$0.03 &     0.00 &     0.08 &     0.00 &  $-$0.01 &  $-$0.04 &     0.01 &     0.07 &     0.00 &     0.05 &  1 \\          
\ion{Ni}{i}    &     0.15 &     0.08 &     0.08 &     0.04 &     0.05 &     0.02 &     0.04 &     0.04 &  $-$0.02 &     0.09 &     0.06 &     0.09 &     0.06 &     0.04 &  6 \\
\ion{Ce}{ii}   &     0.31 &     0.37 &     0.20 &     0.23 &     0.21 &     0.26 &     0.15 &     0.30 &     0.12 &     0.22 &     0.31 &     0.15 &     0.24 &     0.08 &  9 \\
\ion{Nd}{ii}  &          &     0.43 &     0.47 &          &          &     0.39 &     0.12 &          &     0.07 &     0.45 &     0.31 &     0.49 &     0.34 &     0.16 &  1 \\
\ion{Yb}{ii}   &  $-$0.05 &     0.14 &     0.08 &     0.11 &          &  $-$0.04 &     0.13 &     0.13 &     0.01 &     0.22 &     0.05 &     0.00 &     0.07 &     0.08 &  1 \\
\carbiso&       25 &       20 &       25 &       27 &       24 &       28 &       23 &       20 &       25 &       20 &       26 &        6 &       22 &        6 & CO   \\
C       &  $-$0.25 &  $-$0.11 &  $-$0.30 &  $-$0.32 &  $-$0.23 &  $-$0.24 &  $-$0.27 &  $-$0.34 &  $-$0.40 &  $-$0.15 &  $-$0.20 &  $-$0.61 &  $-$0.28 &     0.13 & CO, \ion{C}{i}   \\
N       &     0.54 &     0.46 &     0.56 &     0.61 &     0.45 &     0.48 &     0.44 &     0.60 &     0.45 &     0.51 &     0.41 &     0.67 &     0.52 &     0.08 & CN   \\
O       &     0.02 &     0.07 &     0.03 &     0.09 &  $-$0.01 &     0.07 &  $-$0.12 &  $-$0.07 &  $-$0.14 &     0.04 &  $-$0.04 &  $-$0.11 &  $-$0.01 &     0.08 & OH   \\
\hline                                                                                                                                                          
\multicolumn{12}{l}{$^{*}$ This study.}
\end{tabular}                                                                                                                                                   
\end{table*}

In previous optical studies we have used the Line Depth 
Ratio (LDR) method for estimations of initial \teff\ values.
The LDR method was first developed by \cite{gray91} and later studied by 
several authors such as \cite{biazzo07a,biazzo07b}, whose formulae we 
utilized in Paper~1.
LDR temperatures are calculated from the ratios central depths of 
\teff-sensitive absorption lines to those of lines that are relatively 
insensitive to \teff.
Such line pairs are relatively insensitive
to \logg\ and [Fe/H] for disk-metallicity red giants.
The method is completely free of of photometric uncertainties, interstellar 
reddening, and extinction.
In a recent study, \cite{fukue15} identified 18 absorption lines in the
$H$ band to provide \teff$_{\rm ,LDR}$ calibrations for nine absorption 
line pairs, for mostly G- and K-type giants and supergiants. 
This study provides a new opportunity to estimate effective temperature,
perhaps the most important atmospheric parameter, without having any 
information from the optical spectral range.

We applied the \cite{fukue15} $H$-band LDR method to our IGRINS data, 
measuring line depths of their recommended absorption pairs, and calculated 
\mbox{$T_{\rm eff,LDR}$} of the NGC~6940 RGs (Table~\ref{tab-model}). 
In Figure~\ref{teff}, we plot the optical LDR temperatures (Paper~1) along 
with the infrared LDR temperatures that were obtained using the temperature 
scales reported by \cite{fukue15}, and compare them with the 
spectroscopic temperatures. 
LDR temperatures from the $IR$ region are in accord with 
the optical LDR temperatures, and they agree well with spectroscopic 
\teff\ values for all but the coolest star, MMU~105.
All other program stars are at least 150~K warmer than MMU~105 according
to the Paper~1 spectroscopic analyses, and the $(B-V)_0$ and $(V-K)_0$
values (Table~\ref{tab-basic}) for this star also suggest a lower temperature.
Without more examples like MMU~105, we cannot pursue this point further.
Comparison of the average LDR and spectral temperatures yields,
$\langle$\teff$_{\rm ,LDR}$ $-$ \teff$_{\rm ,spec}\rangle$~=  3~$\pm$~10~K in the 
optical (Paper 1), and
$\langle$\teff$_{\rm ,LDR}$ $-$ \teff$_{\rm ,spec}\rangle$~= 16~$\pm$~14~K in the $IR$.
It is clear that the $H$-band LDR method provides reliable \teff\ value 
independent of other atmospheric parameters, which will 
become especially useful for RGs in other clusters 
that are not observable in the optical spectral region due to large 
interstellar extinction.

\section{Abundances from the Infrared Window}\label{irabs}

We determined the abundances of 20 elements
from the high-resolution $H$- and $K$-band spectra of NGC~6940 RG stars. 
17 of the elements studied here were also investigated in Paper~1.
We followed the analytical methods of \cite{afsar18}, in particular
adopting their line lists of atomic and molecular data without change.
We performed synthetic spectrum analyses to measure the $IR$ abundances.
We measured the abundances of H-burning (C, N, O), $\alpha$ (Mg, Si, S, Ca),
light odd-Z (Na, Al, P, K), Fe-group (Sc, Ti, Cr, Fe, Co, Ni) and \ncap\ 
(Ce, Nd, Yb) elements, as well as \carbiso\ ratios.
We begin with an overview of the results, and follow with details
of the analysis and comparison with the optical abundances in subsections
to follow.

\begin{figure}
 \includegraphics[width=\columnwidth]{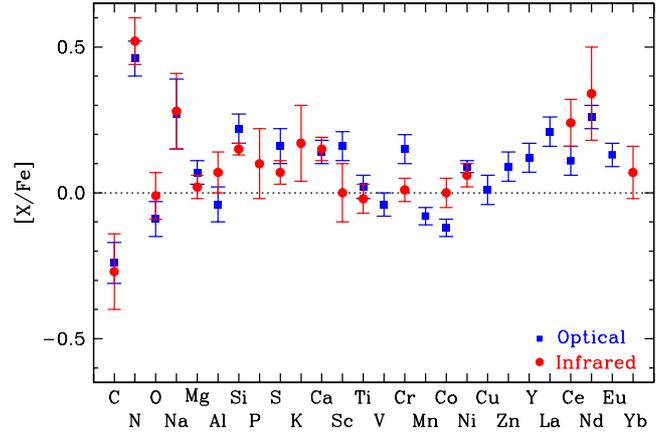}
      \caption{NGC~6940 cluster mean elemental abundances from optical 
                (blue symbols) and $IR$ (red symbols) spectral region.
                The data for this figure are from Table~\ref{tab-abunds}.
                For elements represented by two species (Cr and Ti), 
                the average of the species is displayed.}
     \label{InfOpt}
\end{figure}

In the bottom part of Table~\ref{tab-abunds} we list 
the $IR$-based relative abundance ratios of our 12 NGC~6940 program stars.
Figure~\ref{InfOpt} displays a summary of the mean optical and $IR$ abundances.
Blue squares and red dots are the calculated mean abundances 
from 12 RGs in the optical and $IR$ region, respectively. 
Error bars represent the standard deviations of each element of its
cluster mean abundance.
Inspection of this figure shows that there is general agreement of abundances
derived from the two spectral domains, with abundance differences rarely
exceeding the mutual abundance uncertainties.
Defining $\Delta^{\rm IR}_{\rm opt}$[A/B]~= [A/B]$_{\rm IR}$ $-$ [A/B]$_{\rm opt}$, we find 
$\langle\Delta^{\rm IR}_{\rm opt}$[X/Fe]$\rangle$ = 0.00~$\pm$~0.02 ($\sigma$~= 0.09)
for 17 species with both optical and $IR$ abundances.

\begin{figure*}
  \leavevmode
      \epsfxsize=14cm
      \epsfbox{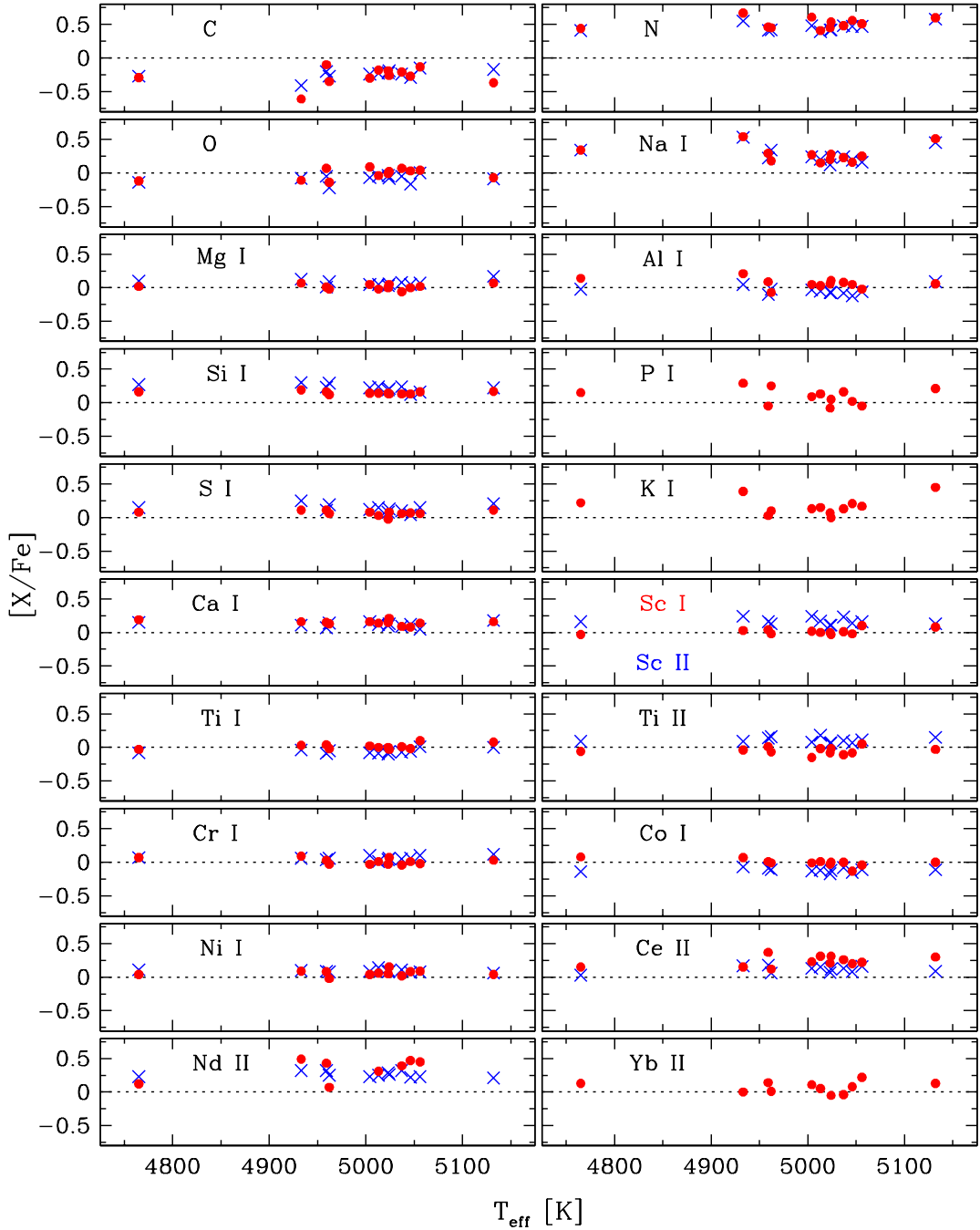}
      \caption{Mean species abundances for all NGC~6940 program stars 
               plotted as functions of their \teff\ values.
               The panels labeled simply C, N, and O are based on multiple
               abundance indicators that are discussed in \S\ref{cnoiso}. Optical
               and $IR$ abundances are shown with blue crosses and red dots,
               respectively. 
               In the Sc panel, \ion{Sc}{i} (red dots) and \ion{Sc}{ii} 
               (blue crosses) represent the measurements from $IR$ and optical, 
               respectively.
 }
      \label{Abd3}
\end{figure*}

In Figure~\ref{Abd3} we show species mean abundances 
for each program star as a function of its effective temperature.
This figure reinforces the optical/$IR$ agreement and the small star-to-star
abundance scatters evident in Figure~\ref{InfOpt}. 
We compare only the neutral species of Cr due to its existence
in both regions.
Additionally, this figure illustrates that there are no significant abundance
drifts with \teff, albeit over the small 4765$-$5132~K range occupied by
the RG stars of our sample.

\begin{figure}
\includegraphics[width=\columnwidth]{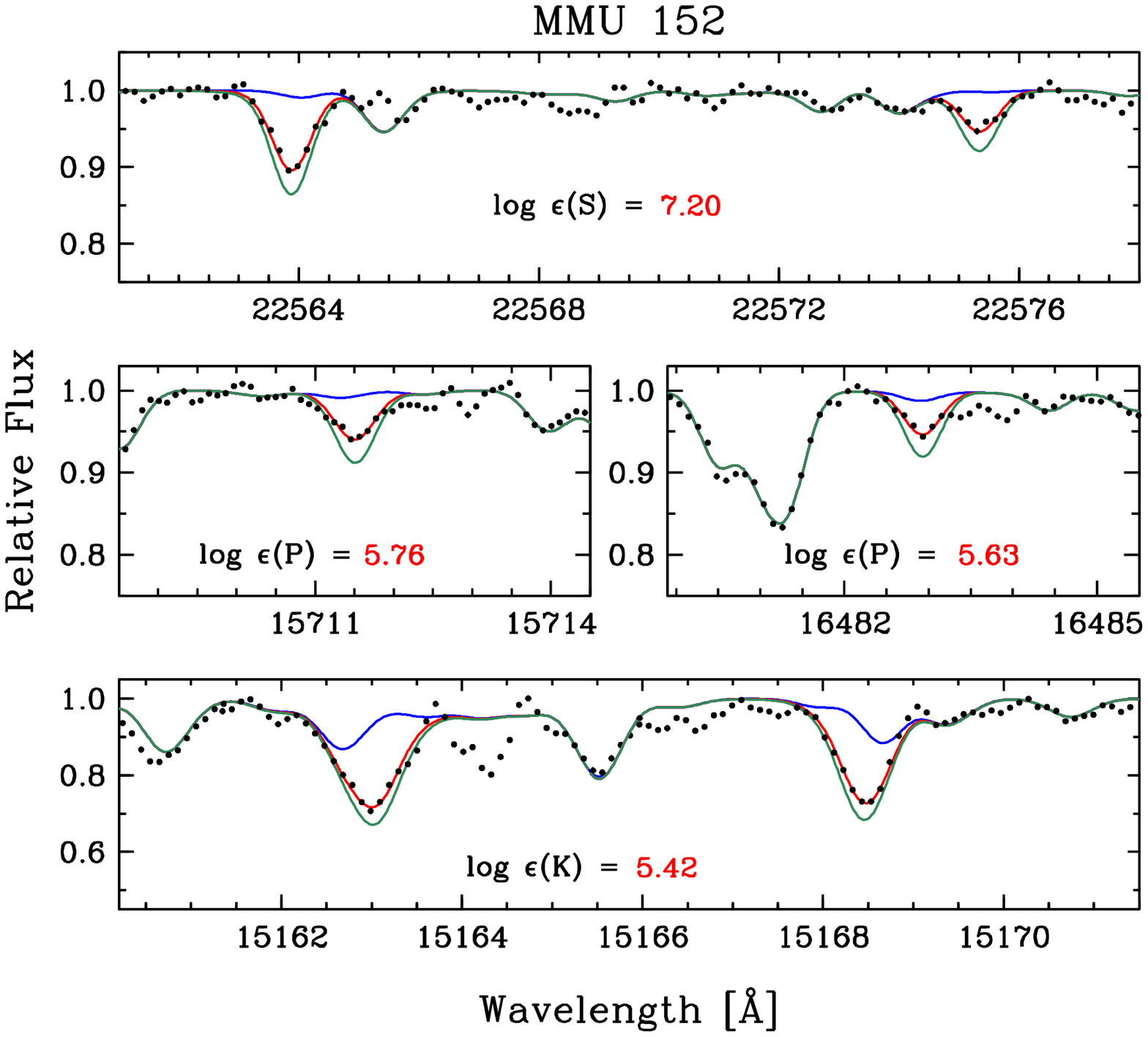}
      \caption{Observed (points) and synthetic spectra 
               (colored lines) of transitions for three light elements 
               rarely studied in the optical wavelength domain.
               In each panel the blue line represents a synthesis with
               no contribution by the element of interest, the red line
               is for the abundance that best matches the observed spectrum,
               and the green line represent the synthesis 
               larger than the best fit by 0.3 dex.
  }
      \label{PSK_new}
\end{figure}

\subsection{Fe-group elements:}\label{fegroup}

The standard metallicity element Fe is represented
only by \ion{Fe}{i} on our IGRINS spectra; it is a struggle to detect
any \ion{Fe}{ii} lines in RG $IR$ spectra.
Additionally, there are not many recent laboratory transition probabilities
reported for \ion{Fe}{i}.
The series of lab studies of this species (\citealt{ruffoni14}, 
\citealt{denhartog14}, \citealt{belmonte17}) have no lines in the $H$ and
$K$ bands.
Therefore our \ion{Fe}{i} $gf$-values were all derived from 
reverse solar analyses by  \cite{afsar18}, and thus our Fe abundances
are differential ones.
They are listed for each star in Table~\ref{tab-model}, from which
we derive an NGC~6940 cluster abundance of 
$\langle$[Fe/H]$\rangle_{\rm IR}$~= 0.02 ($\sigma$~=~0.06).
The means for the optical lines (which employed lab transition probabilities
for both \ion{Fe}{i} and \ion{Fe}{ii} are
$\langle$[Fe/H]$\rangle_{\rm I,opt}$~= $-$0.02 ($\sigma$~=~0.06) and
$\langle$[Fe/H]$\rangle_{\rm II,opt}$~= $-$0.07 ($\sigma$~=~0.06).
These three metallicity estimates are all solar within the uncertainties.

The Fe-group comprises elements Sc through Zn, 
$Z$~=~ 21$-$30.  We detected useful transitions of five Fe-group species 
in the $IR$ (\ion{Sc}{i}, \ion{Ti}{i}, \ion{Ti}{ii}, 
\ion{Cr}{i}, \ion{Co}{i} and \ion{Ni}{i}).
From their mean [X/Fe] values in Table~\ref{tab-abunds} we compute 
$\langle$[X/Fe]$\rangle_{\rm IR}$ = 0.01 ($\sigma$~= 0.04,
6 species).
A caution needs to be made for \ion{Ti}{ii}, which
could be determined only from one feature in the $IR$,
and \ion{Sc}{i} abundances which could be measured only from two weak lines in the $K$ band.
In general, the $IR$ results compare well with the ones from the optical region, in which for
10 elements we find 
$\langle$[X/Fe]$\rangle_{\rm opt}$ = 0.04 ($\sigma$~= 0.11, 11 species).
Both optical and $IR$ Fe-group relative abundances are consistent
with their solar values.

\subsection{Alpha elements}\label{alphas}

The $\alpha$-elements, Mg, Si, Ca and S, each have about
10 absorption neutral-species lines in $H$- and $K$-bands 
(Table~\ref{tab-abunds}). 
Mg, Si and Ca lines are usually stronger compared to those of S.
We illustrate this in the top panel of the Figure~\ref{PSK_new}, showing
observed/synthetic spectrum comparisons of two of the main \ion{S}{i} 
lines in MMU~152. 
Almost all $\alpha$-elements have abundances somewhat above solar. 
The mean of all $\alpha$ group elements for the cluster is
$\langle$[$\alpha$/Fe]$\rangle$$_{\rm IR}$$\equiv$ \begin{scriptsize}$\dfrac{1}
{4}$\end{scriptsize}([Mg/Fe]+[Si/Fe]+[S/Fe]+[Ca/Fe]) = 0.10 ($\sigma$~=~0.06).
This agrees reasonably well with the mean [$\alpha$/Fe] from the optical, 
0.15 ($\sigma$~=~0.06).
We derived the optical Si and Ca abundances in Paper~I from $EW$s of large 
sets of lines, finding very small internal scatter.
However, Mg abundances were derived only from two strong \mbox{Mg\,{\sc i}}
lines at 5528.41 and 5711.08 \AA\ by spectrum synthesis analyses. 
The line-to-line scatter of the Mg abundances from these lines reached up to 
about 0.2~dex in some cases. 
In the $IR$, line-to-line scatter is only $\sim$0.05 dex from 10 
\mbox{Mg\,{\sc i}} lines, indicating that more robust Mg abundance can be 
obtained using the IGRINS data.

\subsection{Odd-Z light elements}\label{oddz}
 
Abundances of four odd-Z light elements Na, Al, P and K
were determined from our $IR$ spectra.
Our LTE analysis (in both regions) yields a significant overabundance for Na.
Giving equal weight to the Na abundance averages from $IR$
and optical domains, $\langle$[Na/Fe]$\rangle$~=~0.28, but the star-to-star 
scatter is large, $\sigma$~=~0.13.
The optical and $IR$ abundances are in good agreement:
$\Delta^{\rm IR}_{\rm opt}$[Na/Fe]~= 0.01 ($\sigma$~=~0.07).
For star MMU~132 the $IR$ Na abundance is 0.16~dex smaller than that from
the optical.
Eliminating that star yields smaller scatter in the
optical/IR abundance comparison:
$\Delta^{\rm IR}_{\rm opt}$[Na/Fe]~= 0.03 ($\sigma$~=~0.05).
This suggests that the star-to-star variation in Na abundances is real.
Note that the scatter is driven in large part by very large abundances
for stars MMU~108 and MMU~152 ([Na/Fe]~$\simeq$~0.5).
As previously discussed in Paper~1 (Section 7.3) 
in detail, and also recently studied by, e.g. \cite{smiljanic18}, 
relatively large turnoff mass (2 M$_{\odot}$) of the cluster could 
be associated with the Na overabundance. 
In \S\ref{mmu152}, we will discuss further the Na 
overabundance issue observed in MMU~152.

In contrast to the scatter for Na, the Al abundances 
in our NGC~6940 giants are consistent with 
[Al/Fe]~$\simeq$~0.07 with little evidence of star-to-star variation 
($\sigma$~=~0.07). 
The optical and $IR$ abundances slightly differ from each other with 
$\Delta^{\rm IR}_{\rm opt}$[Al/Fe]~= 0.11 ($\sigma$~=~0.08).
Once again MMU~152 yields high [Al/Fe] values, about 0.1~dex larger than
the cluster means in the two spectral regions.

Abundances of other two odd-Z light elements P and K have 
been analyzed from two neutral-species lines that are present for each
element in the $H$ band.
\cite{moore66} identified three high-excitation \ion{P}{i} lines 
($\chi$~$\geq$~8~eV) in the optical spectral region, but they could not 
be detected in our NGC~6940 spectra, being either blended or extremely
weak (reduced widths log($RW$)~$\equiv$ log($EW/\lambda$)~$<$~$-$6.0).
The \ion{K}{i} resonance line at 7699.0~\AA\ has been the dominant
source of K abundances in the literature, but in our NGC~6940 giants
the 7699~\AA\ line is on the flat/damping part of the curve-of-growth 
log($RW$)~$>$~$-$4.6 and thus relatively insensitive to abundance variations.
Additionally, the \ion{K}{i} resonance lines are subject to severe
NLTE effects, as discussed in, \eg, \cite{takeda02,takeda09}.
\cite{afsar18} suggests that the NLTE effects might be much less for the
high-excitation \ion{K}{i}, but detailed calculations have not been
published.

The middle panels of Figure~\ref{PSK_new} displays the 
synthetic/observational spectrum matches for the \ion{P}{i} 15711.5 and 
16482.9 \AA\ lines in MMU~152, while the bottom panel shows the 
\ion{K}{i} 15163.1 and 15168.4 \AA\ lines.
The cluster means are $\langle$[P/Fe]$\rangle_{\rm IR}$~= $0.10$ 
($\sigma=0.12$) and $\langle$[K/Fe]$\rangle_{\rm IR}$ = $0.17$ ($\sigma=0.13$).
Abundances MMU~108 and especially MMU~152 again are overabundant with respect
to the cluster means, being K overabundant by about 0.2~dex in MMU~152, and
0.3~dex in MMU~108.

\subsection{Neutron-capture Elements}\label{ncapels}

We detected transitions of \ion{Ce}{ii}, 
\ion{Nd}{ii}, and \ion{Yb}{ii} in our $IR$ spectra, to complement 
the ionized optical region transitions of  Y, La, Nd, and Eu in Paper~1,
and additional Ce transitions presented in this study.
Caution should be used in interpreting the $IR$ Nd and Yb abundances, since
they have been derived from one transition each.
All neutron-capture elements are somewhat overabundant, 
but among the rare-earth elements there is a 
split between those whose origin in solar-system material is attributed
more to slow neutron captures (the \spro) and rapid neutron captures
(the \rpro).  
Using the $r$- and $s$-fractions in Table~10 of \cite{simmerer04},
the ``mostly $s$ elements'' are La (75\% \spro), Ce (81\%), and Nd (58\%),
and the ``mostly $r$ elements'' are Eu (97\% \rpro) and Yb (68\%).
Other $r$/$s$ assessments, \eg, \cite{sneden08} give similar fractions
for these elements.
A simple mean of both $IR$ and optical La, Ce, and Nd abundances is
$\langle$[\spro/Fe]$\rangle$~$\simeq$~0.23, while that for Eu and Yb is
$\langle$[\rpro/Fe]$\rangle$~$\simeq$~0.10.
This is suggestive of a mild \spro\ overabundance in NGC~6940, but the effect
is too weak to be pursued with our spectra.

\subsection{The CNO Group}\label{cnoiso}

\begin{figure}
\includegraphics[width=\columnwidth]{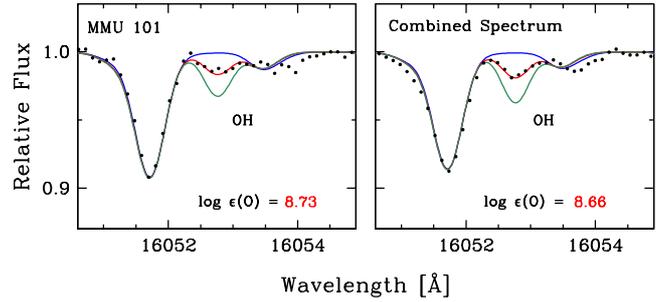}
      \caption{Observed and synthetic spectra of the very 
               weak 16042.8~\AA\ OH (3-1) vibration-rotation band line.
               In the left panel MMU~101 is illustrated; it has one of the 
               best detections of this line.
               In the right panel the mean spectrum of all program stars
               is compared to syntheses computed with an average NGC~6940;
               see the discussion in \S\ref{cnoiso}.
               Synthetic spectra are color-coded as the same way 
               as in Figure~\ref{PSK_new}.
              }
      \label{Oxy_new}
\end{figure}

The $H$ and $K$ spectral region provides many useful 
molecular features such as OH, CN  and CO. 
These transitions can greatly strengthen the reliability of CNO abundances 
and \carbiso\ ratios derived from optical spectroscopy.
We applied iterative spectrum synthesis to these features to derive fresh
CNO abundances for  NGC~6940 RG stars purely from our IGRINS spectra.

Abundances of C and O must be derived together 
iteratively because these elements are linked through CO formation.
However for our NGC~6940 RGs the problem is not severe, because the
strengths of species used to determine O abundances 
([\species{O}{i}] in the optical, OH in the $IR$)
are relatively insensitive to assumed C abundances.  
In Paper~1 we derived cluster means 
$\langle$[C/N]$\rangle$~=~$-$0.56 and $\langle$[C/O]$\rangle$~=~$-$0.17, 
in reasonable agreement with expectations for solar metallicity RG abundance 
ratios resulting from interior CN-cycle hydrogen fusion and envelope mixing.
In absolute abundance terms these ratios are
$\langle$\eps{C/N}$\rangle$~$\simeq$~0.05 and
$\langle$\eps{C/O}$\rangle$~$\simeq$~$-$0.43.
For our NGC\,6940 program giants
the molecular CO association is small, especially
so for O, being about 2.5 times more abundant than C.
Our molecular equilibrium computations yield N(CO)/N(C)~$\simeq$~0.09
and N(CO)/N(O)~$\simeq$~0.03 in atmospheric line-forming layers.
Thus internally consistent C and O abundances could be derived in 1-2 iterations
of the C- and O-containing species.

We began by assuming an average C abundance from Paper~1 
and deriving O abundances from 3-15 OH lines.
Although there are many OH lines especially in the $H$ band, they are usually 
weak (sometimes undetectable) and often are blended in our stars. 
We illustrate this in the left panel of Figure~\ref{Oxy_new} by showing 
observed and synthetic spectra of the 16052~\AA\ line in MMU~101, one of the 
best detections of OH in our program stars.
This line is not obviously contaminated by lines of other 
species, but it is less than 2\% deep.
In order to convince ourselves of the reality of this and other chosen OH lines,
we experimented by combining the spectra of all 12 RGs into one single 
``cluster mean'' spectrum (adopting an average atmospheric parameter set
for the cluster mean: \teff = 4996 K, \logg = 3.0, [M/H] = $-$0.06, \vmicro = 1.17 \kmsec).
All of the lines that we eventually used for individual stars were identified
in the combined spectrum, and many other also.
The 16052~\AA\ line in this spectrum is shown in the right panel of
Figure~\ref{Oxy_new}; the increased $S/N$ of the combined spectrum can be 
easily seen by inspection of right and left panels of the figure.

The O abundance from the combined spectrum (using 21 OH 
lines) of 12 RGs was [O/Fe]~=~0.04 ($\sigma$~= 0.08).
The mean of the individual O abundances from OH lines in 12 RGs is
[O/Fe] = $-$0.01 ($\sigma$~=~0.08, Table~\ref{tab-abunds}).       
This value is consistent with O abundance that derived from the 
[\ion{O}{i}] 6300~\AA\ line,[O/Fe]~=~$-$0.09 ($\sigma$~=~0.06), 
within the mutual uncertainties of both O species.
Both O abundance indicators should be treated with caution.
In addition to the weakness and blending issues of OH lines, this species
makes up less than 1\% of the total O content in atmospheric line-forming
regions; uncertainties in OH syntheses are magnified in the O elemental
abundances derived from OH.
[\ion{O}{i}] lines arise from the dominant O-containing 
species, as discussed above.
But these transitions are plagued with multiple contaminants, both stellar and
telluric, and are difficult to work with in NGC~6940 (see Figure~8 of
Paper~1).
We suggest that the $IR$ OH transitions may yield abundances for this cluster
that are more trustworthy then
those obtained with the optical forbidden lines.

Adopting the O abundance derived from OH features we then 
derived C abundances from the many prominent $K$-band CO molecular features,
specifically those of the \iso{12}{CO} first overtone $\Delta v$~=~2 (2$-$0) 
and (3$-$1) bands.
All CO transitions yielded consistent C abundances with very small internal
scatter, about 0.03~dex. 
The mean NGC~6940 abundance from CO features was 
$\langle$[C/Fe]$\rangle$~=~$-$0.29 ($\sigma=0.14$).
In Paper~1 we obtained the optical C abundances from two C$_{2}$ Swan 
band heads: the (0-0) band head at 5165~\AA, which is heavily blended by 
atomic absorption lines, and the (0-1) band head at 5635~\AA, which is very 
weak and also blended with other features.
Using these two regions we obtained a cluster mean of [C/Fe] = $-0.26$ 
($\sigma=0.08$), consistent with the new CO-based abundance.

In this study, we also investigated the high excitation 
\ion{C}{i} lines for independent C abundance estimations.
This species has detectable transitions over the whole optical and $IR$ 
spectral range. 
In the optical we measured \ion{C}{i} lines at 5380 and 8335~\AA. 
In the $IR$, we were able to locate four useful \ion{C}{i} lines at 
16022, 16890, 17456 and 21023~\AA. 
We summarize the C abundances from \ion{C}{i}, C$_2$ and CO in the optical 
and IR in Table~\ref{tabc}. 
The various features yield consistent C abundances in all NGC 6940 RGs. 
For the cluster, $\langle$[C/Fe]$\rangle$~=~$-$0.24 ($\sigma$~=~0.07) from the 
optical features and $\langle$[C/Fe]$\rangle$~=~$-$0.28 ($\sigma$~=~0.13) 
from the IR features (Table~\ref{tab-abunds}).

\begin{table}
  \caption{[C/Fe] abundances in optical and infrared regions.} 
  \label{tabc}
  \begin{tabular}{@{}lcccccc@{}}
  \hline
Star & \ion{C}{i} & C$_2$ & \ion{C}{i} &   CO & mean & mean \\
     &            opt &   opt &           $IR$ & $IR$ &  opt & $IR$ \\
\hline
MMU 28  &  $-$0.17 &  $-$0.22 &  $-$0.23 &  $-$0.28 &  $-$0.19 &  $-$0.25 \\
MMU 30  &  $-$0.20 &  $-$0.19 &  $-$0.11 &  $-$0.10 &  $-$0.20 &  $-$0.11 \\
MMU 60  &  $-$0.28 &  $-$0.31 &  $-$0.28 &  $-$0.31 &  $-$0.29 &  $-$0.30 \\
MMU 69  &  $-$0.24 &  $-$0.25 &  $-$0.28 &  $-$0.36 &  $-$0.24 &  $-$0.32 \\
MMU 87  &  $-$0.22 &  $-$0.22 &  $-$0.23 &  $-$0.23 &  $-$0.22 &  $-$0.23 \\
MMU 101 &  $-$0.22 &  $-$0.26 &  $-$0.28 &  $-$0.19 &  $-$0.24 &  $-$0.24 \\
MMU 105 &  $-$0.27 &  $-$0.28 &  $-$0.25 &  $-$0.29 &  $-$0.27 &  $-$0.27 \\
MMU 108 &  $-$0.14 &  $-$0.21 &  $-$0.33 &  $-$0.34 &  $-$0.17 &  $-$0.34 \\
MMU 132 &  $-$0.20 &  $-$0.34 &  $-$0.37 &  $-$0.43 &  $-$0.27 &  $-$0.40 \\
MMU 138 &  $-$0.15 &  $-$0.16 &  $-$0.13 &  $-$0.16 &  $-$0.15 &  $-$0.15 \\
MMU 139 &  $-$0.22 &  $-$0.23 &  $-$0.22 &  $-$0.20 &  $-$0.23 &  $-$0.21 \\
MMU 152 &  $-$0.37 &  $-$0.44 &  $-$0.61 &  $-$0.62 &  $-$0.41 &  $-$0.61 \\
\hline
\end{tabular}
\end{table}

With the newly-derived O and C abundances in the IR, we 
calculated N abundances from $\sim$18 selected CN molecular lines 
located between 15000 and 15500~\AA.  
The cluster mean N abundance for 12 RGs is $\langle$[N/Fe]$\rangle$~=~0.52 
($\sigma$~=~0.08, Table~\ref{tab-abunds}), in good accord with optical mean 
for the cluster, $\langle$[N/Fe]$\rangle$~=~0.46 ($\sigma$~=~0.06).

\begin{figure}
\includegraphics[width=\columnwidth]{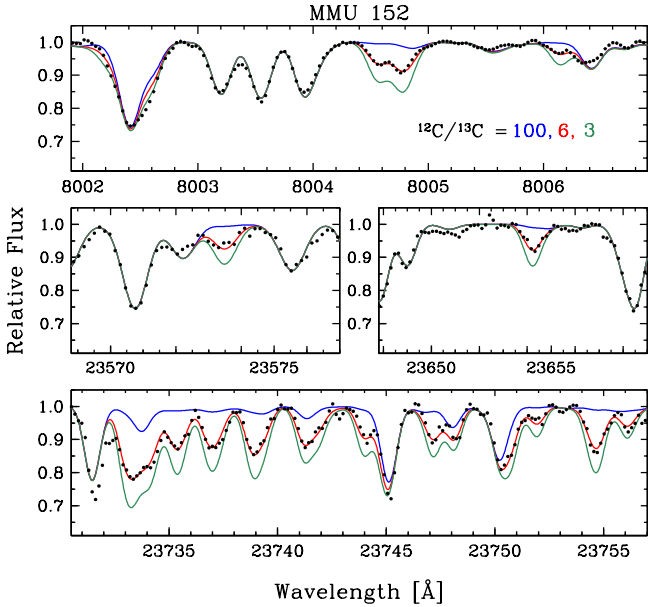}
      \caption{Observed and synthetic spectra illustrating the very low 
               carbon isotopic ratio of NGC~6940 MMU~152.
               In each panel the blue, red, and green lines represent
               \carbiso~=100, 6, and 3, respectively.
               The top panel is centered on the triplet or \iso{13}{CN}
               red system (2-0) lines that have been the primary indicators
               of the isotopic ratio in optical studies of RG stars.
               The middle panels show representative \iso{13}{CO} (2-0) 
               R-branch vibration-rotation band lines, and the bottom
               panel shows the \iso{13}{CO} (3-1) R-branch bandhead.}
      \label{Ciso_152}
\end{figure}

The first overtone \iso{12}{CO} ($\Delta$v~=~2) band 
heads (2$-$0) and (3$-$1) are accompanied by the \iso{13}{CO} band heads near 
23440 and 23730~\AA, respectively. 
These features allowed us to determine more robust \carbiso\ values for
the cluster members compared to the ones obtained in Paper~1 only from the 
\iso{13}{CN} feature near 8003~\AA. 
In Figure~\ref{Ciso_152}, we plot both regions for MMU~152,
the RG member with the lowest \carbiso\ in our sample.
Generally the CN and CO isotopic indicators yield consistent mean values 
leading to cluster mean $\langle$\carbiso$\rangle$~=~20, with a 
few stars (notably MMU 69 and MMU 101) indicating much larger isotopic 
ratios from IR CO than from optical CN features. 
We list the \carbiso\ ratios of all the RGs for both spectral 
region in Table~\ref{tab-iso}.

\begin{table}
  \caption{Carbon isotropic ratios of optical and infrared regions.} 
  \label{tab-iso}
  \begin{tabular}{@{}lccc@{}}
  \hline
 Stars   &  $^{13}$CN  &  $^{13}$CO  &  $^{13}$CO \\
           & (8004 \AA) &  (23440 \AA)  &  (23730 \AA)    \\         
 \hline
MMU 28	&	25	&	    	&	25	\\
MMU 30	&	15	&	15	&	25	\\
MMU 60	&	20	&	    	&	25	\\
MMU 69	&	10	&	27	&	27	\\
MMU 87	&	25	&	18	&	30	\\
MMU 101	&	15	&	30	&	25	\\
MMU 105	&	15	&	20	&	25	\\
MMU 108	&	20	&	20	&	20	\\
MMU 132	&	25	&	25	&	25	\\
MMU 138	&	20	&	18	&	22	\\
MMU 139	&	15	&	25	&	27	\\
MMU 152	&	6	&	6	&	6	\\
 \hline
\end{tabular}
\end{table}

Taking NGC~6940 RG optical and $IR$ CNO indicators as a whole,
we find [C/Fe]~=~$-$0.26, [N/Fe]~=~0.49, [O/Fe]~=~$-$~0.05,
and \carbiso~=~20.
Excluding MMU~152 from the sample the cluster RGs have [C/Fe]~=~$-$0.24,
[N/Fe]~=~0.48, [O/Fe]~=~$-$~0.05, and \carbiso~$\simeq$~22.
The \cite{asplund09} initial C and N abundances are \eps{C}~=~8.43,
\eps{N}~=~7.83, or \eps{C/N}$_{\rm init}$~=~0.60.
RG observed mean abundance ratio [C/N]~=~$-$0.75 becomes 
\eps{C/N}$_{\rm RG}$~=~$-$0.15.
This value is somewhat lower than
the original stellar evolution predictions of 
\cite{iben64,iben67a,iben67b}, as discussed by \cite{lambert81}.
The predictions given in Table~6 of \citeauthor{lambert81} are for assumed
initial ratios \eps{C/N}~= 0.86, 0.68, and 0.50.
The corresponding RG values are \eps{C/N}$_{\rm pred}$~= 0.20, 0.13, and 0.05.
The 2~M$_\odot$ RGs in NGC6940 with \citeauthor{asplund09} initial abundances
would be predicted to have \eps{C/N}$_{\rm pred}$~$\simeq$~0.1. 
The predicted isotopic ratios from \citeauthor{lambert81} are nearly identical
to our RG observed mean value.  
For initial values \carbiso$_{\rm init}$~=~89, 50, and 25, they predict that
\carbiso$_{\rm RG}$~=~21, 18, and 13.
For the reasonable (unprovable) assumption that \carbiso$_{\rm init}$~$\simeq$~90,
the predicted RG value of 21 is nearly identical to that of our NGC~6940 
program stars.

\subsection{Hydrogen Fluoride}\label{hftext}

Inspection of our $K$-band spectra revealed no obvious
stellar absorption at 23358.3~\AA, the wavelength of the single unblended
HF line in this spectral region.
The spectra of two of the stars, MMU 105 and MMU 152, were examined 
in greater detail for the presence of the HF feature.
The details of HF analyses to obtain an abundance of fluorine
are described in \cite{pilachowski15}.
                                                              
Synthetic spectra of the HF region were computed using MOOG and 
the model atmospheres determined for each star.               
An excitation potential of $\chi$~=~0.227~eV was adopted from HITRAN        
molecular line database \citep{rothman13}.                    
The oscillator strength log~$gf$~=~$-$3.971 was adopted from  
\cite{lucatello11}, and the dissociation energy used by MOOG, 
D$_0$~=5.8698~eV, is consistent with the \cite{jonsson14} calculations.     
We adopted wavelengths, excitation potentials, and gf-values from 
\cite{goorvitch94} for spectrum synthesis of the neighboring CO (2-0)       
and (3-1) vibration-rotation lines.                           
For the handful of atomic lines in the spectrum, we adopted line 
parameters from the Vienna Atomic Line Database (\citealt{kupka00} and      
references therein) for our spectrum synthesis.               
                                                              
\begin{figure}
\includegraphics[width=\columnwidth]{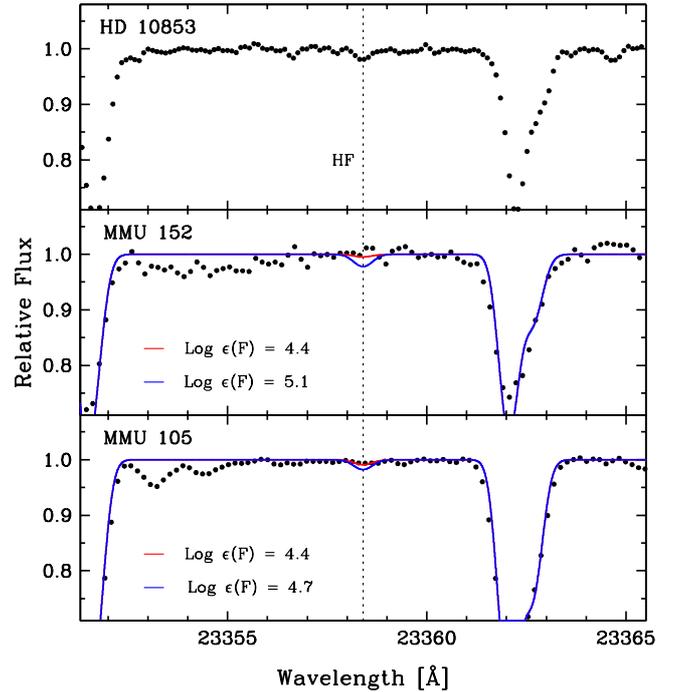}
      \caption{Spectra of two giants, MMU 105 and MMU 152, along a 
               comparison star HD 10853.  
               Synthetic spectra are shown at the solar abundance of
               F (red) and an upper limits on the stellar abundance of F 
               (blue) for each NGC 6940 star.  
               The position of the HF feature is marked with a dashed 
               vertical line.}
      \label{hffig}
\end{figure}

The synthetic spectra, calculated with a fluorine abundance   
\eps F~=~4.40 \citep{maiorca14}, are compared to the observed stellar        
spectra in Figure~\ref{hffig}.                                
For comparison, we include the slightly cooler K3.5 V star HD 10853         
from \cite{pilachowski15}.                                    
They found that for solar metallicity stars, HF is not reliably
present in stars warmer than $\approx$~4500 K.                 
\citeauthor{pilachowski15} adopted atmospheric parameters        
(\teff~=~4600~K, \logg~=~4.65, [Fe/H]~=~$-$0.12)             
for HD~10853 from \cite{mishenina08}, and determined          
a fluorine abundance of \eps F~=~4.27 for the star.          
                                                              
For MMU 105 at \teff~=~4762~K, we find an upper limit to the abundance      
of fluorine to be \eps F~$<$~4.7 from synthetic spectra.     
For MMU 152, the upper limit is higher, \eps F~$<$~5.1 due to the lower    
$S/N$ ratio for this spectrum.                                
Both stars are consistent with a solar abundance of fluorine, although we   
cannot rule out a slight enhancement. 

While we cannot rule out a small excess in the abundance of 
fluorine, an enhancement of F through mass transfer from a carbon-enhanced 
asymptotic giant branch (AGB) star is unlikely.  
AGB stars are thought to be significant producers of fluorine 
(\citealt{jorissen92}, \citealt{abia09}).
Redetermination of the fluorine abundances in peculiar red giants 
\citep{abia15} suggests that large fluorine excesses ([F/H]~$>$~+0.3 or 
\eps F~$>$~4.6) are limited primarily to stars with C/O ratios near 
unity (C/O~$<$~1.08).
Stars with C/O greater than this value have near-solar abundances of 
fluorine (\eps F~$\simeq$~4.3). 

\section{Stellar Evolution Modelling of NGC\,6940}
\label{isoch}

Interpretation of the light element abundances in our
NGC\,6940 giants requires knowledge of their evolutionary state(s).
Therefore we modelled the observed photometric properties of the 
stars in NGC\,6940 with both the MESA \citep{paxton:11,paxton:13} and 
Victoria \citep{vandenberg:14} stellar evolutionary codes.  
The former code was used to compute the evolutionary tracks of stars having 
the main-sequence turnoff (MSTO) mass from the zero-age main sequence (ZAMS) 
to the core helium-burning phase on the assumption of the same input physics 
and control parameters as described by \cite{denissenkov:17}.
We assumed [Fe/H]~=~0.0 (Paper~I) in these computations, and an initial solar 
abundance mix recommended by \cite{asplund09}.
An important component of these calculations was the treatment of interior
convection.
Stars with larger masses than the Sun are predicted to have convective cores 
during their MS phase.
It has long been known that some degree of core overshooting 
must also occur in order to satisfy such empirical constraints as the 
properties of eclipsing binaries and the morphology of the MS blue hook in
the color-magnitude diagrams (CMDs) of young ($\lta 4$ Gyr) open clusters, 
which occurs when the central H abundance is exhausted. 
Since the pioneering work that established the importance of convective 
core overshooting in intermediate-mass and massive stars 
\citep[e.g.][]{bressan81,bertelli86,maeder87}, there have been many studies devoted to 
the calibration of the extent of overshooting as a function of mass and metallicity 
\citep[see, e.g.][]{vandenberg:06,bressan12,stancliffe15,deheuvels16,claret18}.
All stellar models must implement some treatment of convective core 
overshooting in order to match photometric observations.
In the next subsection we describe our treatment of convective overshoot
as it applies to stars similar in mass to NGC\,6940.

\subsection{Treatment of Interior Convection}\label{convect}
\label{isoch}

As in the case of convective helium cores, convective overshooting outside the
Schwarzschild boundary of a convective H core during the MS evolution has been
modelled with the exponentially decaying diffusion coefficient
\bea
D_\mathrm{ov}(r) = D_\mathrm{conv}(r_0)\exp\left[\frac{-2(r-r_0)}{f_\mathrm{ov}H_P}\right],
\label{eq:Dov}
\eea
as proposed by \cite{herwig:97} based on the hydrodynamical simulations of
\cite{freytag:96}.  
Here, $D_\mathrm{conv}(r_0)$ is the convective diffusion coefficient both 
inside the convective core and close to its boundary calculated with the 
mixing length theory, $H_P$ is the pressure scale height, and $f_\mathrm{ov}$ 
is a free parameter whose value must be constrained empirically.
\cite{herwig:00} found that the value $f_\mathrm{ov}\approx 0.016$ enabled him
to reproduce the earlier models of \cite{schaller:92} that had used a step-like
prescription for convective overshooting; i.e., the convective boundary was
arbitrarily relocated to a distance $\alpha_\mathrm{ov}H_P$
outside the Schwarzschild boundary.  
Their models were able to successfully match the upper envelope of the MS 
(the so-called ``terminal-age main sequence" or TAMS) defined by 65 star 
clusters and associations of different ages when the value of the convective 
overshooting parameter $\alpha_\mathrm{ov} = 0.2$ was used.

Victoria models \citep[see][]{vandenberg:06} are similar except that the 
enlargement of a central convective core is derived from a parameterized 
form of the integral equations derived by \cite{roxburgh:89} for the maximum 
possible size of such a core.  
By introducing a parameter $F_\mathrm{over}$ into the integral equations,
where $0 \le F_\mathrm{over} \le 1$, and then determining best estimates of its
value from inter-comparisons of synthetic and observed CMDs for clusters spanning
a wide range in age, as well as observations of selected binary stars,
VandenBerg et al.~were able to deduce how $F_\mathrm{over}$ varies with mass.
According to their Fig.~1, $F_\mathrm{over}$ appears to ramp up from a value 
near 0.0 at the lowest mass with sustained core convection to a value of
$\approx 0.55$ at a higher mass by about 0.5 M$_\odot$, and to remain 
essentially constant with further increases in mass.  
Lacking any evidence to the contrary, \citeauthor{vandenberg:06} assumed 
that this variation is nearly independent of metallicity, aside from a 
zero-point adjustment.  
That is, the functional relationship between $F_\mathrm{over}$ and mass is 
taken to be the same as the one that they derived for solar abundances, 
except that, for other abundance choices, it is shifted in mass until
$F_\mathrm{over} = 0.0$ coincides with the mass at which the transition is made 
between stars that have radiative cores at the end of the MS phase to those 
that retain convective cores until central H exhaustion.

One of the totally eclipsing binary stars considered by \cite{vandenberg:06} 
was TZ For, which has close to the solar metallicity (like NGC\,6940) and whose
components have masses very close to the MSTO mass of NGC\,6940.  
Indeed, this binary provides a unique benchmark test of the efficiency of 
convective overshooting in the turnoff stars of NGC\,6940, especially as 
\cite{gallenne:15} have recently determined to very high precision 
($\sim 1\%$) the effective temperatures, luminosities, masses, and radii of its components.  
Follow-up studies by \cite{valle:17} and \cite{higl:18} 
examined the implications of the improved stellar parameters of this binary 
for the calibration of core overshooting in their stellar models.\footnote{
Both of these studies assumed that the secondary of TZ For has 
$T_\mathrm{eff} = 6650 \pm 200$~K, which corresponds to the adopted 
temperature of the model atmosphere that was used in the spectroscopic analysis 
carried out by \cite{gallenne:15}.
However, the latter gave $6350 \pm 70$~K as their best estimate of its 
temperature from an analysis of the interferometric $H$-band flux ratio.  
The higher temperature, by 300~K, used by \citeauthor{valle:17} and by 
\citeauthor{higl:18} has the consequence that their 
values of the extent of core overshooting would have been underestimated somewhat.}

\begin{figure}
\includegraphics[width=\columnwidth]{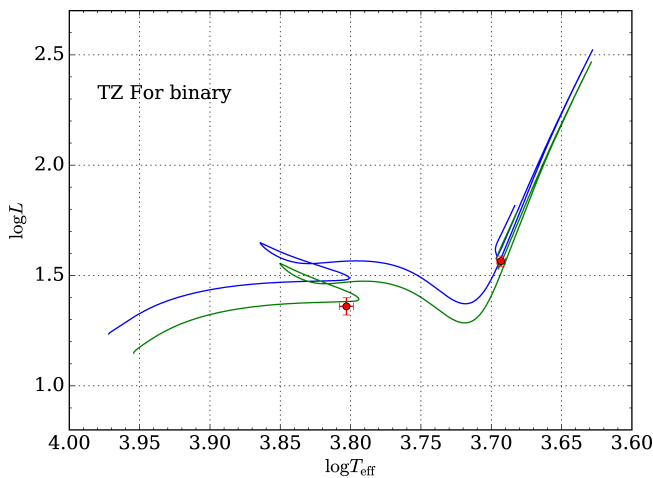}
\caption{The positions of the components of the eclipsing binary system 
         TZ Fornacis on the HRD (red dots with errorbars) and their fits 
         with the MESA stellar evolution models computed using the parameters
         adopted for these stars by \protect\cite{gallenne:15}. 
         The best fits with the relative age difference between the stars 
         less than $\sim 1\%$ are obtained assuming that
         [Fe/H]\,$=0.05$, $Y=0.25$, and $f_\mathrm{ov} = 0.035$.
        }
\label{TZFor}
\end{figure}

Our simulations with the MESA code of the evolution of the TZ For components
with the parameters adopted for them by \cite{gallenne:15} are able to match
their positions on the H-R diagram with a relative age difference of less than
1\% only if we assume that $f_\mathrm{ov} = 0.035$ and the primary component is
in the core He-burning stage of evolution (see Figure~\ref{TZFor}).  
Very similar conclusions were reached by \cite{higl:18}, however see 
\cite{constantino:18}.  
We have therefore used this value of the convective overshooting parameter 
in generating the MESA models that have been applied to NGC\,6940.  
It turns out that an equally good fit to the properties of the secondary 
component, which provides the main constraint on core overshooting in 
$\sim$ 2~M$_\odot$ stars having close to the solar metallicity,
can be obtained using models that are generated by the Victoria code assuming
the abundances of the metals given by \cite{asplund09} and the overshooting
prescription described above.

\subsection{Evolutionary Tracks and Isochrones for NGC\,6940}\label{ios6940}

The Victoria code has state-of-the-art model interpolation 
codes \citep{vandenberg:14}, so we used it to produce the 1.15 Gyr,
2.0 M$_\odot$ turnoff mass
isochrone that provides the best fit to the CMD of NGC\,6940 (the black 
curve in Figure~\ref{NGC6940GaiaDR2}).
The grid of models employed for this interpolation assume solar abundances,
a helium abundance of $Y=0.27$, and a range in mass from 1.0 to 2.0 M$_\odot$. 
The blue curve in Figure~\ref{NGC6940GaiaDR2} is the MESA evolutionary track 
computed for M = 2.0 M$_\odot$ and the same initial chemical composition 
as that used to generate the isochrone. 
The isochrone and the evolutionary tracks have been transposed 
to the observed plane using the reddening-corrected bolometric corrections 
given by \cite{cas18} for the $Gaia$ G$_{\rm BP}$, G, and G$_{\rm RP}$ bandpasses.
On the assumption of a true distance modulus, (m-M)$_{\rm 0}$ = 10.11, 
the isochrone provides the best overall fit to the cluster photometry if 
$E(B-V) = 0.21$.  With this choice, the isochrone appears to be slightly too 
red relative to the upper MS stars, while deviating somewhat to the blue 
of the MS stars at G $>$ 16 (not shown).

It is possible that an improved fit could be obtained by 
adopting slightly different cluster parameters.  As noted in \S\ref{obs}, 
the cluster parallax has a $1\sigma$ uncertainty of 0.052 $mas$, 
which corresponds to a distance modulus uncertainty of about $\pm 0.12$ mag.  
Furthermore, \cite{lindegren18} have suggested that current Gaia 
parallaxes suffer from a zero-point error of $\sim$0.03 $mas$, in the sense 
of being too small.  Insofar as the foreground reddening is concerned, 
the 3D dust maps provided by \cite{green18,green15} yield 
$E(B-V) = 0.24 - 0.25$, with an uncertainty of $\pm$~0.02$-$0.03, depending 
on the adopted distance of NGC 6940.  However, even if the actual cluster 
properties (including the age and the turnoff mass) differ somewhat from 
those that we have adopted or derived, the predicted chemical properties 
of the RC stars, which are of particular importance for the present 
investigation, would not change significantly.

The close similarity between the blue track and the isochrone 
at G$_{\rm BP}$$-$G$_{\rm RP}$  $>$ 1.2, when the predicted mass along the isochrone 
is close to 2.0 M$_\odot$ and the variation in mass has become very small, 
illustrates the good agreement
between our MESA and Victoria models; see \cite{denissenkov:17} for a more
complete discussion of the indistinguishability of  
MESA and Victoria models when their input physical ingredients are similar.

The colors of the red-giant branches (RGBs) of both the blue track and the
isochrone are $\sim$0.05 magnitude redder than they should be in order to 
fit the CMD positions of the NGC\,6940 red giants. 
Similar color offsets have been found in the fits of isochrones to the 
CMDs of other open clusters \citep[e.g.][]{choi16,hidalgo18}
and globular clusters \citep[e.g.][]{vandenberg:13,fu18}.
They are likely caused by deficiencies in the treatment of convection, the 
adopted atmospheric boundary condition, and/or the assumed 
color--$T_\mathrm{eff}$~relations.
Recent work aimed at calibrating the mixing-length 
theory using 3D hydrodynamic atmosphere models \citep[and references 
therein]{magic:15} indicates that the mixing-length parameter should decrease 
towards higher $T_\mathrm{eff}$, lower surface gravity, and higher 
metallicity, but the consequences for stellar models appear to be small 
\citep{mosumgaard:18}.
As an illustration, we computed a track with the same parameters as for the 
blue one of Figure~\ref{NGC6940GaiaDR2}, but with the
mixing length increased by 10\%; see the red curve in this figure.
It provides quite a good fit to the observed giants, though one could 
potentially accomplish the same thing by altering the boundary condition for 
the pressure at $T=T_\mathrm{eff}$ by assuming a different 
$T$--$\tau$~relation or by adopting the photospheric pressures from a grid 
of model atmospheres; see, e.g.,
\cite{vandenberg:08}; \cite{vandenberg:14}, their Fig.~15; \cite{salaris:15}.
What is much more important is that both the red and blue tracks have the 
same minimum luminosity during core He-burning, and that this luminosity
agrees very well with the observed magnitudes of the red giants in NGC\,6940.
This is the reason why the luminosity of the red clump is a good distance 
indicator \citep[see, e.g.][]{stanek97,nataf13}.

\begin{figure*}
\begin{center}
\includegraphics[width=15cm]{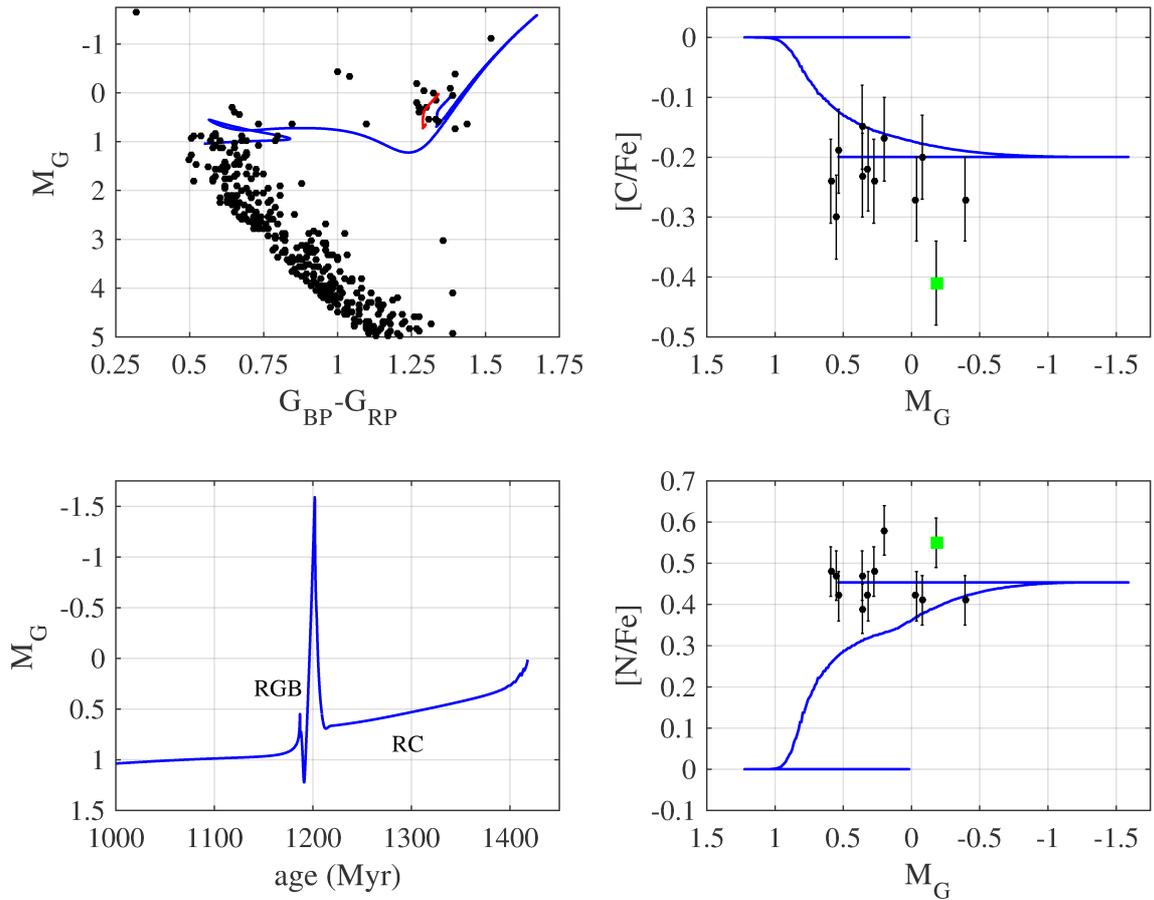}
\caption{Top-left panel: The CMD of NGC\,6940 with the blue track and a 
         red-clump fragment of the red track from Figure~\ref{NGC6940GaiaDR2}.
         Right panels: the corresponding changes of the surface C and 
         N abundances compared with the ones observed in the red-clump stars.
         Bottom-left panel: the RGB (for the absolute 
         magnitude M$_{\rm G}$ decreasing from $\sim$1.0 to $\sim$$-$1.5) 
         and red-clump (horizontal-branch) evolutionary timescales. 
         MMU 152 is illustrated with a green square.
        }
\label{ngc6940Results}
\end{center}
\end{figure*}

Figure~\ref{ngc6940Results} shows a superposition of the blue track onto
the cluster CMD (top-left panel), the predicted evolutionary timescales for
the RGB and red-clump (horizontal branch) phases (bottom-left panel), and the
variations of the surface abundances of C and N (right-hand panels). 
Note, in particular, that the lifetime on the RGB, where M$_{\rm G}$ 
decreases from $\sim$1.0 to $\sim-$1.5,
is much shorter than the evolutionary timescale of the
red clump.  Therefore, most, if not all, of the red-giant stars in NGC\,6940 are
predicted to be core He-burning red clump stars. Furthermore, because the
H-burning shell in our RGB models (along both the blue and red tracks) never
reaches and erases the mean molecular weight barrier left behind by the bottom
of the convective envelope at the end of the first dredge-up (FDU), the
first-ascent red giants are not expected to experience extra mixing along the
RGB.  Nor will they experience a helium core flash; they will instead ignite
core helium quiescently.  These are among the expected consequences of
significant core overshooting during the MS stage.  Thus, the surface chemical
composition of the red-clump stars in NGC\,6940 should reflect only the changes
that occurred during the FDU. 
Our computations support this conclusion for all of the 
red-clump stars except MMU\,152, the star with the lowest [C/Fe] ratio
in the Figure~\ref{ngc6940Results} top-right panel, illustrated with a green
square.

\section{MMU~152 AND ITS ANOMALOUS LIGHT ELEMENT ABUNDANCES}\label{mmu152}

MMU~152 has anomalous light element abundances compared the 11 program stars:
it has a much lower C, much higher N and Na, and  somewhat higher Al, P and K.
Additionally MMU~152 has a very low carbon isotopic ratio, 
\carbiso~=~6~$\pm$~1.
This is easily seen in both optical CN and $IR$ CO features 
(Figure~\ref{Ciso_152}).
Normal high metallicity disk field and open cluster red 
giants do not exhibit such unusual light element abundances.
These MMU~152 abundances resemble more what is found in globular cluster
evolved stars.  

Large star-to-star and cluster-to-cluster correlated and anti-correlated 
variations are found among the CNONaMgAl elements 
(\eg, \citealt{gratton12,carretta18,bastian18} and references therein).
Briefly, abundances of C and N are anti-correlated as qualitatively expected 
from normal CN cycle hydrogen fusion.
But the magnitude of C deficiencies and N enhancements are often much more
extreme than seen in disk RG stars.
More importantly there are well-documented O-N, O-Na, often O-Al, and 
sometimes Mg-Al anti-correlations seen in globular cluster stars.

The interior syntheses leading to these correlated light 
element abundances is most easily understood as various high temperature 
(T~$\geq$~4$\times$10$^7$~K) proton-capture sequences, mainly the O-N, 
Ne-Na, and Mg-Al cycles.
The observed Mg-Al anti-correlations in some globular clusters such as M13 
\citep{kraft97,carretta09a,carretta09b} require H-fusion temperatures 
T~$\geq$~7$\times$10$^7$~K.
It is believed that most low-mass globular cluster the stars with anomalous 
light element abundances must have acquired them at or near birth from the 
ejecta of intermediate-mass stars within the clusters.
Those stars would have had the requisite high-temperature interior layers 
capable of generating the requisite O-N, Ne-Na, and Mg-Al cycles.

The light element abundance distribution of NGC~6940 MMU~152 
resembles those of some RGs in metal-rich globular clusters.
Averaging the optical and $IR$ abundances in Table~\ref{tab-abunds}, MMU~152
has [C/Fe]~=~$-$0.51, [N/Fe]~=~0.61, [O/Fe]~=~$-$0.10, [Na/Fe]~=~0.53,
[Mg/Fe]~=~0.10, [Al/Fe]~=~0.13, and \carbiso~=~6.
These abundances are like those found in the disk
globular cluster 47~Tuc (\eg, \citealt{brown89}, \citealt{shetrone03},
\citealt{carretta09a,carretta09b}, 
\citealt{johnson14}, \citealt{thygesan14}).
The maximum relative O abundances of 47~Tuc giants, [O/Fe]~$\simeq$~0.5,
are larger than those of any NGC~6940 program star, but globular clusters
have high initial O abundances compared to OCs.

Recently, \cite{pancino18} has reported discovery of 
rapidly-rotating dwarf stars in several OCs that have relatively low O and 
high Na abundances, sometimes accompanied by low Mg abundances.
Since OC's, including NGC~6940, show no indications of multiple stellar
generations, \citeauthor{pancino18} suggests that the causes of these main 
sequence anomalies are internal, most likely from a combination of several
mechanisms such as diffusion, mixing, rotation, and mass loss.
MMU\,152 might be an elderly red-giant version of such a rapidly
rotating dwarf star, but we have no observational way to confirm this idea.

If the core H-fusion regions of MMU~152 were hot enough to ignite the O-N 
and Ne-Na cycles, then either there was more vigorous envelope mixing
or larger mass loss than occurred in other RGs of this cluster.  
If the slightly anomalous radial velocity of MMU~152 is due to the presence
of an unseen binary companion, then that presumably higher-mass star could 
have given the unique CNO abundances and \carbiso\ to MMU~152 through 
mass transfer.  
The ($V-K$) value for MMU~152 is not unusual compared to other RGs
(Table~\ref{tab-motions}), thus not hinting at any extended dust envelope
for this star.

Alternatively, MMU\,152 may be a counterpart of the Li-rich red-clump stars 
\citep{smiljanic:18} that also show evidence of enhanced extra mixing.
The Li enrichment in this case could be a transient event 
\citep{denissenkov:04} that the star might have experienced in the past. 
At its present age the original Li has been destroyed, so that only the 
changes in the surface abundances of C and N are left as signatures of 
this enhanced mixing.  
There are various speculations regarding the cause of the enhanced extra 
mixing \citep[e.g.][]{kumar:18}. 
In the case of MMU\,152 we can rule out mixing caused by the core He flash
because, according to our simulations, He ignited quiescently in its core.
Rotational mixing during the long-lived red-clump phase could be responsible,
but this hypothesis has yet to be supported by stellar evolution simulations.

Our C, N, and \carbiso\ abundance predications are similar to those of 
\cite{charbonnel:10} for a solar-metallicity 2 M$_\odot$ star.
They used a smaller amount of convective overshooting,
corresponding to $\alpha_\mathrm{ov}=0.1$.  When we reduce $f_\mathrm{ov}$
to a value of 0.014 \citep{constantino:18}, the H-burning shell in our RGB 
models does erase the mean molecular weight barrier, and core He is ignited 
in the form of a flash.  
However, we do not expect any significant contribution of RGB
extra mixing to the change of the surface chemical composition because of
its short timescale and expected shallow depth. 
This is confirmed by results in Table 2 of \citeauthor{charbonnel:10}.
Unlike us, they have also taken into account rotational mixing, showing 
that it can slightly reduce
the carbon isotopic ratio and significantly lower the surface Li abundance.
Therefore, the observed $^{12}$C/$^{13}$C isotopic ratio of 10--15 in some of
the red-clump stars in NGC\,6940 could be a result of rotational mixing on the
MS.  The difficulty with this possibility is that the Li abundance should be
very low as well, which is not the case, at least for the stars MMU 69 and 139.

Finally, considering the fluorine abundances presented in \S\ref{hftext},
any excess of fluorine in the atmosphere of MMU\,152 would most likely
be accompanied by an excess of carbon.  
The observed deficiency of C in MMU 152's atmosphere suggests that any 
mass transfer onto the star from an AGB companion must have occurred
before any carbon dredge-up from He burning, and prior to any fluorine
production.
Together with the low carbon isotope ratio (\carbiso~$\simeq$~6), 
the low carbon abundance and unenhanced fluorine abundance suggests 
that any mass transfer onto MMU 152 would likely come from an early AGB star.
Although our RV measurements do not show any evidence
of binarity for MMU~152, the possibility of being a double star with 
large separation based on its position in the colour-magnitude diagram 
has been discussed by \cite{mermilliod89}.

\section{Conclusions}\label{cocl}

We have observed 12 red giant members of NGC~6940 with
the high-resolution $H$- and $K$-band spectrograph IGRINS.
Temperature-sensitive line-depth ratios derived from our $IR$ spectra
are in good accord with those in the optical, and yield \teff\ values
consistent with those derived from standard line analyses done in Paper~1.

Adopting the atmospheric parameters derived from Paper~1, 
we have derived abundances of 19 elements from 20 species measured on the 
IGRINS spectra.  
For elements in common with those studied in Paper~1 we find good relative
abundance agreement.
Abundances of elements P, K, Ce, and Yb are reported for the first time 
in NGC~6940.
For some species we believe that the $IR$ abundances are more reliable than
those determined from optical transitions.
The availability of multiple species in the optical and $IR$ strengthens 
the abundances of both C (\ion{C}{i}, C$_2$, and CO) and O
([\ion{O}{i}], OH).

Having the advantage of $Gaia$ DR2, we used up-to-date
kinematic parameters and determined the most probable members of the NGC~6940. 
We then applied Victoria isochrones and MESA models 
to the updated CMD of the cluster and investigated 
the evolutionary status of our targets, paying special attention to those light elements
whose abundances can be altered by interior fusion cycles and envelope
mixing episodes.
The isochrones suggest our RGs as core He-burning red clump 
stars with mostly undergone canonical FDU mixing and the CNO abundances of
our targets are consistent with these standard stellar evolutionary predictions.

The only exception in our sample comes with the RG member
MMU 152, which has the lowest [C/Fe] and \carbiso\ values among 
our sample. The low \carbiso\ ratio of MMU 152 suggests extra mixing 
to be involved during its evolution, which could be the result
of rotational mixing that took place on the MS phase. The effect of rotational mixing 
also could not be ruled for the members with \carbiso$\sim$10$-$15.
Besides the extreme depletion of C, MMU 152 exhibits 
larger enhancements of N and Na, and slight enhancement of Al.
This star's surface light element abundances show signs of high-temperature 
H burning.
Whether these abundances have been encouraged by severe envelope mass loss
or by more efficient mixing is not clear.

We will continue searching for other cluster RGs with peculiar CNO
abundances in future IGRINS studies of M67, NGC~752, and several 
dust-obscured open clusters for which optical high-resolution spectroscopy
is impractical.

\section*{Acknowledgments}

We thank the anonymous referee for her/his comments and suggestions that improved
the quality of the paper.
We thank Karin Lind and Henrique Reggiani
for helpful discussions on this work.
Our work has been supported by The Scientific and Technological 
Research Council of Turkey (T\"{U}B\.{I}TAK, project No. 116F407), by the US
National Science Foundation (NSF, grant AST 16-16040),
and by the University of Texas Rex G. Baker, Jr. Centennial Research Endowment.
This work used the Immersion Grating Infrared Spectrometer (IGRINS) that 
was developed under a collaboration between the University of Texas at 
Austin and the Korea Astronomy and Space Science Institute (KASI) with 
the financial support of the US National Science Foundation under grant 
AST-1229522, of the University of Texas at Austin, and of the Korean GMT 
Project of KASI.
This work has made use of data from the European Space Agency (ESA) mission
$Gaia$ ($\rm https://www.cosmos.esa.int/gaia$), processed by the {\it Gaia}
Data Processing and Analysis Consortium (DPAC,
$\rm https://www.cosmos.esa.int/web/gaia/dpac/consortium$). 
Funding for the DPAC has been provided by national institutions, in 
particular the institutions participating in the $Gaia$ Multilateral Agreement.
This research has made use of NASA's Astrophysics Data System 
Bibliographic Services; the SIMBAD database and the VizieR service, 
both operated at CDS, Strasbourg, France. 
This research has made use of the WEBDA database, operated at the 
Department of Theoretical Physics and Astrophysics of the Masaryk University,
and the VALD database, operated at Uppsala University, the 
Institute of Astronomy RAS in Moscow, and the University of Vienna.

This paper includes data taken at The McDonald Observatory of The 
University of Texas at Austin.

  \bibliographystyle{mnras}{}
  \bibliography{totbib}

\begin{thebibliography}{}
\makeatletter
\relax
\def\mn@urlcharsother{\let\do\@makeother \do\$\do\&\do\#\do\^\do\_\do\%\do\~}
\def\mn@doi{\begingroup\mn@urlcharsother \@ifnextchar [ {\mn@doi@}
  {\mn@doi@[]}}
\def\mn@doi@[#1]#2{\def\@tempa{#1}\ifx\@tempa\@empty \href
  {http://dx.doi.org/#2} {doi:#2}\else \href {http://dx.doi.org/#2} {#1}\fi
  \endgroup}
\def\mn@eprint#1#2{\mn@eprint@#1:#2::\@nil}
\def\mn@eprint@arXiv#1{\href {http://arxiv.org/abs/#1} {{\tt arXiv:#1}}}
\def\mn@eprint@dblp#1{\href {http://dblp.uni-trier.de/rec/bibtex/#1.xml}
  {dblp:#1}}
\def\mn@eprint@#1:#2:#3:#4\@nil{\def\@tempa {#1}\def\@tempb {#2}\def\@tempc
  {#3}\ifx \@tempc \@empty \let \@tempc \@tempb \let \@tempb \@tempa \fi \ifx
  \@tempb \@empty \def\@tempb {arXiv}\fi \@ifundefined
  {mn@eprint@\@tempb}{\@tempb:\@tempc}{\expandafter \expandafter \csname
  mn@eprint@\@tempb\endcsname \expandafter{\@tempc}}}

\bibitem[\protect\citeauthoryear{{Abia}, {Recio-Blanco}, {de Laverny},
  {Cristallo}, {Dom{\'{\i}}nguez}  \& {Straniero}}{{Abia}
  et~al.}{2009}]{abia09}
{Abia} C.,  {Recio-Blanco} A.,  {de Laverny} P.,  {Cristallo} S.,
  {Dom{\'{\i}}nguez} I.,   {Straniero} O.,  2009, \mn@doi [ApJ]
  {10.1088/0004-637X/694/2/971}, \href
  {http://adsabs.harvard.edu/abs/2009ApJ...694..971A} {694, 971}

\bibitem[\protect\citeauthoryear{{Abia}, {Cunha}, {Cristallo}  \& {de
  Laverny}}{{Abia} et~al.}{2015}]{abia15}
{Abia} C.,  {Cunha} K.,  {Cristallo} S.,   {de Laverny} P.,  2015, \mn@doi
  [A\&A] {10.1051/0004-6361/201526586}, \href
  {http://adsabs.harvard.edu/abs/2015A%26A...581A..88A} {581, A88}

\bibitem[\protect\citeauthoryear{{Af{\c s}ar} et~al.,}{{Af{\c s}ar}
  et~al.}{2016}]{afsar16}
{Af{\c s}ar} M.,  et~al., 2016, \mn@doi [ApJ] {10.3847/0004-637X/819/2/103},
  \href {http://adsabs.harvard.edu/abs/2016ApJ...819..103A} {819, 103}

\bibitem[\protect\citeauthoryear{{Af{\c s}ar} et~al.,}{{Af{\c s}ar}
  et~al.}{2018}]{afsar18}
{Af{\c s}ar} M.,  et~al., 2018, \mn@doi [ApJ] {10.3847/1538-4357/aada0c}, \href
  {http://adsabs.harvard.edu/abs/2018ApJ...865...44A} {865, 44}

\bibitem[\protect\citeauthoryear{{Arenou} et~al.,}{{Arenou}
  et~al.}{2018}]{aren18}
{Arenou} F.,  et~al., 2018, \mn@doi [A\&A] {10.1051/0004-6361/201833234}, \href
  {http://adsabs.harvard.edu/abs/2018A%26A...616A..17A} {616, A17}

\bibitem[\protect\citeauthoryear{{Asplund}, {Grevesse}, {Sauval}  \&
  {Scott}}{{Asplund} et~al.}{2009}]{asplund09}
{Asplund} M.,  {Grevesse} N.,  {Sauval} A.~J.,   {Scott} P.,  2009, \mn@doi
  [ARA\&A] {10.1146/annurev.astro.46.060407.145222}, \href
  {http://adsabs.harvard.edu/abs/2009ARA%26A..47..481A} {47, 481}

\bibitem[\protect\citeauthoryear{{Bastian} \& {Lardo}}{{Bastian} \&
  {Lardo}}{2018}]{bastian18}
{Bastian} N.,  {Lardo} C.,  2018, \mn@doi [ARA\&A]
  {10.1146/annurev-astro-081817-051839}, \href
  {http://adsabs.harvard.edu/abs/2018ARA%26A..56...83B} {56, 83}

\bibitem[\protect\citeauthoryear{{Belmonte}, {Pickering}, {Ruffoni}, {Den
  Hartog}, {Lawler}, {Guzman}  \& {Heiter}}{{Belmonte}
  et~al.}{2017}]{belmonte17}
{Belmonte} M.~T.,  {Pickering} J.~C.,  {Ruffoni} M.~P.,  {Den Hartog} E.~A.,
  {Lawler} J.~E.,  {Guzman} A.,   {Heiter} U.,  2017, \mn@doi [ApJ]
  {10.3847/1538-4357/aa8cd3}, \href
  {http://adsabs.harvard.edu/abs/2017ApJ...848..125B} {848, 125}

\bibitem[\protect\citeauthoryear{{Bertelli Motta} et~al.,}{{Bertelli Motta}
  et~al.}{2018}]{bertelli18}
{Bertelli Motta} C.,  et~al., 2018, \mn@doi [MNRAS] {10.1093/mnras/sty1011},
  \href {http://adsabs.harvard.edu/abs/2018MNRAS.478..425B} {478, 425}

\bibitem[\protect\citeauthoryear{{Bertelli}, {Bressan}, {Chiosi}  \&
  {Angerer}}{{Bertelli} et~al.}{1986}]{bertelli86}
{Bertelli} G.,  {Bressan} A.,  {Chiosi} C.,   {Angerer} K.,  1986, A\&ASS,
  \href {http://adsabs.harvard.edu/abs/1986A%26AS...66..191B} {66, 191}

\bibitem[\protect\citeauthoryear{{Bharat Kumar}, {Singh}, {Eswar Reddy}  \&
  {Zhao}}{{Bharat Kumar} et~al.}{2018}]{kumar:18}
{Bharat Kumar} Y.,  {Singh} R.,  {Eswar Reddy} B.,   {Zhao} G.,  2018, \mn@doi
  [ApJL] {10.3847/2041-8213/aac16f}, \href
  {http://adsabs.harvard.edu/abs/2018ApJ...858L..22B} {858, L22}

\bibitem[\protect\citeauthoryear{{Biazzo}, {Frasca}, {Catalano}  \&
  {Marilli}}{{Biazzo} et~al.}{2007a}]{biazzo07a}
{Biazzo} K.,  {Frasca} A.,  {Catalano} S.,   {Marilli} E.,  2007a, \mn@doi
  [Astronomische Nachrichten] {10.1002/asna.200710781}, \href
  {http://adsabs.harvard.edu/abs/2007AN....328..938B} {328, 938}

\bibitem[\protect\citeauthoryear{{Biazzo} et~al.,}{{Biazzo}
  et~al.}{2007b}]{biazzo07b}
{Biazzo} K.,  et~al., 2007b, \mn@doi [A\&A] {10.1051/0004-6361:20077374}, \href
  {http://adsabs.harvard.edu/abs/2007A%26A...475..981B} {475, 981}

\bibitem[\protect\citeauthoryear{{B{\"o}cek Topcu}, {Af{\c s}ar}  \&
  {Sneden}}{{B{\"o}cek Topcu} et~al.}{2016}]{bocek16}
{B{\"o}cek Topcu} G.,  {Af{\c s}ar} M.,   {Sneden} C.,  2016, \mn@doi [MNRAS]
  {10.1093/mnras/stw1974}, \href
  {http://adsabs.harvard.edu/abs/2016MNRAS.463..580B} {463, 580}

\bibitem[\protect\citeauthoryear{{Bressan}, {Chiosi}  \& {Bertelli}}{{Bressan}
  et~al.}{1981}]{bressan81}
{Bressan} A.~G.,  {Chiosi} C.,   {Bertelli} G.,  1981, A\&A, \href
  {http://adsabs.harvard.edu/abs/1981A%26A...102...25B} {102, 25}

\bibitem[\protect\citeauthoryear{{Bressan}, {Marigo}, {Girardi}, {Salasnich},
  {Dal Cero}, {Rubele}  \& {Nanni}}{{Bressan} et~al.}{2012}]{bressan12}
{Bressan} A.,  {Marigo} P.,  {Girardi} L.,  {Salasnich} B.,  {Dal Cero} C.,
  {Rubele} S.,   {Nanni} A.,  2012, \mn@doi [MNRAS]
  {10.1111/j.1365-2966.2012.21948.x}, \href
  {http://adsabs.harvard.edu/abs/2012MNRAS.427..127B} {427, 127}

\bibitem[\protect\citeauthoryear{{Brown} \& {Wallerstein}}{{Brown} \&
  {Wallerstein}}{1989}]{brown89}
{Brown} J.~A.,  {Wallerstein} G.,  1989, \mn@doi [AJ] {10.1086/115248}, \href
  {http://adsabs.harvard.edu/abs/1989AJ.....98.1643B} {98, 1643}

\bibitem[\protect\citeauthoryear{{Cantat-Gaudin} et~al.,}{{Cantat-Gaudin}
  et~al.}{2018}]{cantat18}
{Cantat-Gaudin} T.,  et~al., 2018, \mn@doi [A\&A]
  {10.1051/0004-6361/201833476}, \href
  {http://adsabs.harvard.edu/abs/2018A%26A...618A..93C} {618, A93}

\bibitem[\protect\citeauthoryear{{Carretta} et~al.,}{{Carretta}
  et~al.}{2009a}]{carretta09a}
{Carretta} E.,  et~al., 2009a, \mn@doi [A\&A] {10.1051/0004-6361/200912096},
  \href {http://adsabs.harvard.edu/abs/2009A%26A...505..117C} {505, 117}

\bibitem[\protect\citeauthoryear{{Carretta}, {Bragaglia}, {Gratton}  \&
  {Lucatello}}{{Carretta} et~al.}{2009b}]{carretta09b}
{Carretta} E.,  {Bragaglia} A.,  {Gratton} R.,   {Lucatello} S.,  2009b,
  \mn@doi [A\&A] {10.1051/0004-6361/200912097}, \href
  {http://adsabs.harvard.edu/abs/2009A%26A...505..139C} {505, 139}

\bibitem[\protect\citeauthoryear{{Carretta}, {Bragaglia}, {Lucatello},
  {Gratton}, {D'Orazi}  \& {Sollima}}{{Carretta} et~al.}{2018}]{carretta18}
{Carretta} E.,  {Bragaglia} A.,  {Lucatello} S.,  {Gratton} R.~G.,  {D'Orazi}
  V.,   {Sollima} A.,  2018, \mn@doi [A\&A] {10.1051/0004-6361/201732324},
  \href {http://adsabs.harvard.edu/abs/2018A%26A...615A..17C} {615, A17}

\bibitem[\protect\citeauthoryear{{Casagrande} \& {VandenBerg}}{{Casagrande} \&
  {VandenBerg}}{2018}]{cas18}
{Casagrande} L.,  {VandenBerg} D.~A.,  2018, \mn@doi [MNRAS]
  {10.1093/mnrasl/sly104}, \href
  {http://adsabs.harvard.edu/abs/2018MNRAS.479L.102C} {479, L102}

\bibitem[\protect\citeauthoryear{{Charbonnel} \& {Lagarde}}{{Charbonnel} \&
  {Lagarde}}{2010}]{charbonnel:10}
{Charbonnel} C.,  {Lagarde} N.,  2010, \mn@doi [A\&A]
  {10.1051/0004-6361/201014432}, \href
  {http://adsabs.harvard.edu/abs/2010A%26A...522A..10C} {522, A10}

\bibitem[\protect\citeauthoryear{{Choi}, {Dotter}, {Conroy}, {Cantiello},
  {Paxton}  \& {Johnson}}{{Choi} et~al.}{2016}]{choi16}
{Choi} J.,  {Dotter} A.,  {Conroy} C.,  {Cantiello} M.,  {Paxton} B.,
  {Johnson} B.~D.,  2016, \mn@doi [ApJ] {10.3847/0004-637X/823/2/102}, \href
  {http://adsabs.harvard.edu/abs/2016ApJ...823..102C} {823, 102}

\bibitem[\protect\citeauthoryear{{Claret} \& {Torres}}{{Claret} \&
  {Torres}}{2018}]{claret18}
{Claret} A.,  {Torres} G.,  2018, \mn@doi [ApJ] {10.3847/1538-4357/aabd35},
  \href {http://adsabs.harvard.edu/abs/2018ApJ...859..100C} {859, 100}

\bibitem[\protect\citeauthoryear{{Constantino} \& {Baraffe}}{{Constantino} \&
  {Baraffe}}{2018}]{constantino:18}
{Constantino} T.,  {Baraffe} I.,  2018, \mn@doi [A\&A]
  {10.1051/0004-6361/201833568}, \href
  {http://adsabs.harvard.edu/abs/2018A%26A...618A.177C} {618, A177}

\bibitem[\protect\citeauthoryear{{Cunha} et~al.,}{{Cunha}
  et~al.}{2015}]{cunha15}
{Cunha} K.,  et~al., 2015, \mn@doi [ApJL] {10.1088/2041-8205/798/2/L41}, \href
  {http://adsabs.harvard.edu/abs/2015ApJ...798L..41C} {798, L41}

\bibitem[\protect\citeauthoryear{{Cutri} et~al.,}{{Cutri} et~al.}{2003}]{2MASS}
{Cutri} R.~M.,  et~al., 2003, VizieR Online Data Catalog, \href
  {http://cdsads.u-strasbg.fr/abs/2003yCat.2246....0C} {2246, 0}

\bibitem[\protect\citeauthoryear{{Deheuvels}, {Brand{\~a}o}, {Silva Aguirre},
  {Ballot}, {Michel}, {Cunha}, {Lebreton}  \& {Appourchaux}}{{Deheuvels}
  et~al.}{2016}]{deheuvels16}
{Deheuvels} S.,  {Brand{\~a}o} I.,  {Silva Aguirre} V.,  {Ballot} J.,  {Michel}
  E.,  {Cunha} M.~S.,  {Lebreton} Y.,   {Appourchaux} T.,  2016, \mn@doi [A\&A]
  {10.1051/0004-6361/201527967}, \href
  {http://adsabs.harvard.edu/abs/2016A%26A...589A..93D} {589, A93}

\bibitem[\protect\citeauthoryear{{Den Hartog}, {Ruffoni}, {Lawler},
  {Pickering}, {Lind}  \& {Brewer}}{{Den Hartog} et~al.}{2014}]{denhartog14}
{Den Hartog} E.~A.,  {Ruffoni} M.~P.,  {Lawler} J.~E.,  {Pickering} J.~C.,
  {Lind} K.,   {Brewer} N.~R.,  2014, \mn@doi [ApJS]
  {10.1088/0067-0049/215/2/23}, \href
  {http://adsabs.harvard.edu/abs/2014ApJS..215...23D} {215, 23}

\bibitem[\protect\citeauthoryear{{Denissenkov} \& {Herwig}}{{Denissenkov} \&
  {Herwig}}{2004}]{denissenkov:04}
{Denissenkov} P.~A.,  {Herwig} F.,  2004, \mn@doi [ApJ] {10.1086/422575}, \href
  {http://adsabs.harvard.edu/abs/2004ApJ...612.1081D} {612, 1081}

\bibitem[\protect\citeauthoryear{{Denissenkov}, {VandenBerg}, {Kopacki}  \&
  {Ferguson}}{{Denissenkov} et~al.}{2017}]{denissenkov:17}
{Denissenkov} P.~A.,  {VandenBerg} D.~A.,  {Kopacki} G.,   {Ferguson} J.~W.,
  2017, \mn@doi [ApJ] {10.3847/1538-4357/aa92c9}, \href
  {http://adsabs.harvard.edu/abs/2017ApJ...849..159D} {849, 159}

\bibitem[\protect\citeauthoryear{{Freytag}, {Ludwig}  \& {Steffen}}{{Freytag}
  et~al.}{1996}]{freytag:96}
{Freytag} B.,  {Ludwig} H.-G.,   {Steffen} M.,  1996, A\&A, \href
  {http://adsabs.harvard.edu/abs/1996A%26A...313..497F} {313, 497}

\bibitem[\protect\citeauthoryear{{Fu}, {Bressan}, {Marigo}, {Girardi},
  {Montalb{\'a}n}, {Chen}  \& {Nanni}}{{Fu} et~al.}{2018}]{fu18}
{Fu} X.,  {Bressan} A.,  {Marigo} P.,  {Girardi} L.,  {Montalb{\'a}n} J.,
  {Chen} Y.,   {Nanni} A.,  2018, \mn@doi [MNRAS] {10.1093/mnras/sty235}, \href
  {http://adsabs.harvard.edu/abs/2018MNRAS.476..496F} {476, 496}

\bibitem[\protect\citeauthoryear{{Fukue} et~al.,}{{Fukue}
  et~al.}{2015}]{fukue15}
{Fukue} K.,  et~al., 2015, \mn@doi [ApJ] {10.1088/0004-637X/812/1/64}, \href
  {http://adsabs.harvard.edu/abs/2015ApJ...812...64F} {812, 64}

\bibitem[\protect\citeauthoryear{{Gaia Collaboration} et~al.,}{{Gaia
  Collaboration} et~al.}{2016}]{GAIA16}
{Gaia Collaboration} et~al., 2016, \mn@doi [A\&A]
  {10.1051/0004-6361/201629272}, \href
  {http://adsabs.harvard.edu/abs/2016A%26A...595A...1G} {595, A1}

\bibitem[\protect\citeauthoryear{{Gaia Collaboration}, {Brown}, {Vallenari},
  {Prusti}, {de Bruijne}, {Babusiaux}  \& {Bailer-Jones}}{{Gaia Collaboration}
  et~al.}{2018}]{GAIA18b}
{Gaia Collaboration} {Brown} A.~G.~A.,  {Vallenari} A.,  {Prusti} T.,  {de
  Bruijne} J.~H.~J.,  {Babusiaux} C.,   {Bailer-Jones} C.~A.~L.,  2018,
  preprint, \href {http://adsabs.harvard.edu/abs/2018arXiv180409365G} {}
  (\mn@eprint {arXiv} {1804.09365})

\bibitem[\protect\citeauthoryear{{Gallenne} et~al.,}{{Gallenne}
  et~al.}{2016}]{gallenne:15}
{Gallenne} A.,  et~al., 2016, \mn@doi [A\&A] {10.1051/0004-6361/201526764},
  \href {http://adsabs.harvard.edu/abs/2016A%26A...586A..35G} {586, A35}

\bibitem[\protect\citeauthoryear{{Gao} et~al.,}{{Gao} et~al.}{2018}]{gao18}
{Gao} X.,  et~al., 2018, \mn@doi [MNRAS] {10.1093/mnras/sty2414}, \href
  {http://adsabs.harvard.edu/abs/2018MNRAS.481.2666G} {481, 2666}

\bibitem[\protect\citeauthoryear{{Goorvitch}}{{Goorvitch}}{1994}]{goorvitch94}
{Goorvitch} D.,  1994, \mn@doi [ApJS] {10.1086/192110}, \href
  {http://adsabs.harvard.edu/abs/1994ApJS...95..535G} {95, 535}

\bibitem[\protect\citeauthoryear{{Gratton} et~al.,}{{Gratton}
  et~al.}{2012}]{gratton12}
{Gratton} R.~G.,  et~al., 2012, \mn@doi [A\&A] {10.1051/0004-6361/201118491},
  \href {http://adsabs.harvard.edu/abs/2012A%26A...539A..19G} {539, A19}

\bibitem[\protect\citeauthoryear{{Gray} \& {Johanson}}{{Gray} \&
  {Johanson}}{1991}]{gray91}
{Gray} D.~F.,  {Johanson} H.~L.,  1991, \mn@doi [PASP] {10.1086/132839}, 103,
  439

\bibitem[\protect\citeauthoryear{{Green} et~al.,}{{Green}
  et~al.}{2015}]{green15}
{Green} G.~M.,  et~al., 2015, \mn@doi [ApJ] {10.1088/0004-637X/810/1/25}, \href
  {http://adsabs.harvard.edu/abs/2015ApJ...810...25G} {810, 25}

\bibitem[\protect\citeauthoryear{{Green} et~al.,}{{Green}
  et~al.}{2018}]{green18}
{Green} G.~M.,  et~al., 2018, \mn@doi [MNRAS] {10.1093/mnras/sty1008}, \href
  {http://adsabs.harvard.edu/abs/2018MNRAS.478..651G} {478, 651}

\bibitem[\protect\citeauthoryear{{Herwig}}{{Herwig}}{2000}]{herwig:00}
{Herwig} F.,  2000, A\&A, \href
  {http://adsabs.harvard.edu/abs/2000A%26A...360..952H} {360, 952}

\bibitem[\protect\citeauthoryear{{Herwig}, {Bloecker}, {Schoenberner}  \& {El
  Eid}}{{Herwig} et~al.}{1997}]{herwig:97}
{Herwig} F.,  {Bloecker} T.,  {Schoenberner} D.,   {El Eid} M.,  1997, A\&A,
  \href {http://adsabs.harvard.edu/abs/1997A%26A...324L..81H} {324, L81}

\bibitem[\protect\citeauthoryear{{Hidalgo} et~al.,}{{Hidalgo}
  et~al.}{2018}]{hidalgo18}
{Hidalgo} S.~L.,  et~al., 2018, \mn@doi [ApJ] {10.3847/1538-4357/aab158}, \href
  {http://adsabs.harvard.edu/abs/2018ApJ...856..125H} {856, 125}

\bibitem[\protect\citeauthoryear{{Higl}, {Siess}, {Weiss}  \& {Ritter}}{{Higl}
  et~al.}{2018}]{higl:18}
{Higl} J.,  {Siess} L.,  {Weiss} A.,   {Ritter} H.,  2018, \mn@doi [A\&A]
  {10.1051/0004-6361/201833112}, \href
  {http://adsabs.harvard.edu/abs/2018A%26A...617A..36H} {617, A36}

\bibitem[\protect\citeauthoryear{{Hoag}, {Johnson}, {Iriarte}, {Mitchell},
  {Hallam}  \& {Sharpless}}{{Hoag} et~al.}{1961}]{hoag61}
{Hoag} A.~A.,  {Johnson} H.~L.,  {Iriarte} B.,  {Mitchell} R.~I.,  {Hallam}
  K.~L.,   {Sharpless} S.,  1961, Publications of the U.S.~Naval Observatory
  Second Series, \href {http://adsabs.harvard.edu/abs/1961PUSNO..17..343H} {17,
  344}

\bibitem[\protect\citeauthoryear{{Iben}}{{Iben}}{1964}]{iben64}
{Iben} Jr. I.,  1964, \mn@doi [ApJ] {10.1086/148077}, \href
  {http://adsabs.harvard.edu/abs/1964ApJ...140.1631I} {140, 1631}

\bibitem[\protect\citeauthoryear{{Iben}}{{Iben}}{1967a}]{iben67a}
{Iben} Jr. I.,  1967a, \mn@doi [ApJ] {10.1086/149040}, \href
  {http://adsabs.harvard.edu/abs/1967ApJ...147..624I} {147, 624}

\bibitem[\protect\citeauthoryear{{Iben}}{{Iben}}{1967b}]{iben67b}
{Iben} Jr. I.,  1967b, \mn@doi [ApJ] {10.1086/149041}, \href
  {http://adsabs.harvard.edu/abs/1967ApJ...147..650I} {147, 650}

\bibitem[\protect\citeauthoryear{{Johnson} et~al.,}{{Johnson}
  et~al.}{2015}]{johnson14}
{Johnson} C.~I.,  et~al., 2015, \mn@doi [AJ] {10.1088/0004-6256/149/2/71},
  \href {http://adsabs.harvard.edu/abs/2015AJ....149...71J} {149, 71}

\bibitem[\protect\citeauthoryear{{J{\"o}nsson} et~al.,}{{J{\"o}nsson}
  et~al.}{2014}]{jonsson14}
{J{\"o}nsson} H.,  et~al., 2014, \mn@doi [A\&A] {10.1051/0004-6361/201423597},
  \href {http://adsabs.harvard.edu/abs/2014A%26A...564A.122J} {564, A122}

\bibitem[\protect\citeauthoryear{{Jorissen}, {Smith}  \& {Lambert}}{{Jorissen}
  et~al.}{1992}]{jorissen92}
{Jorissen} A.,  {Smith} V.~V.,   {Lambert} D.~L.,  1992, A\&A, \href
  {http://adsabs.harvard.edu/abs/1992A%26A...261..164J} {261, 164}

\bibitem[\protect\citeauthoryear{{Kaeufl} et~al.,}{{Kaeufl}
  et~al.}{2004}]{kaeufl04}
{Kaeufl} H.-U.,  et~al., 2004, in {Moorwood} A.~F.~M.,  {Iye} M.,  eds,
  Society of Photo-Optical Instrumentation Engineers (SPIE) Conference Series
  Vol. 5492, Ground-based Instrumentation for Astronomy. pp 1218--1227,
  \mn@doi{10.1117/12.551480}

\bibitem[\protect\citeauthoryear{{Kharchenko}, {Piskunov}, {R{\"o}ser},
  {Schilbach}  \& {Scholz}}{{Kharchenko} et~al.}{2005}]{khar05}
{Kharchenko} N.~V.,  {Piskunov} A.~E.,  {R{\"o}ser} S.,  {Schilbach} E.,
  {Scholz} R.-D.,  2005, \mn@doi [A\&A] {10.1051/0004-6361:20042523}, \href
  {http://adsabs.harvard.edu/abs/2005A%26A...438.1163K} {438, 1163}

\bibitem[\protect\citeauthoryear{{Kraft}, {Sneden}, {Smith}, {Shetrone},
  {Langer}  \& {Pilachowski}}{{Kraft} et~al.}{1997}]{kraft97}
{Kraft} R.~P.,  {Sneden} C.,  {Smith} G.~H.,  {Shetrone} M.~D.,  {Langer}
  G.~E.,   {Pilachowski} C.~A.,  1997, \mn@doi [AJ] {10.1086/118251}, \href
  {http://adsabs.harvard.edu/abs/1997AJ....113..279K} {113, 279}

\bibitem[\protect\citeauthoryear{{Kupka}, {Ryabchikova}, {Piskunov}, {Stempels}
   \& {Weiss}}{{Kupka} et~al.}{2000}]{kupka00}
{Kupka} F.~G.,  {Ryabchikova} T.~A.,  {Piskunov} N.~E.,  {Stempels} H.~C.,
  {Weiss} W.~W.,  2000, \mn@doi [Baltic Astronomy] {10.1515/astro-2000-0420},
  \href {http://adsabs.harvard.edu/abs/2000BaltA...9..590K} {9, 590}

\bibitem[\protect\citeauthoryear{{Lamb}, {Venn}, {Shetrone}, {Sakari}  \&
  {Pritzl}}{{Lamb} et~al.}{2015}]{lamb15}
{Lamb} M.~P.,  {Venn} K.~A.,  {Shetrone} M.~D.,  {Sakari} C.~M.,   {Pritzl}
  B.~J.,  2015, \mn@doi [MNRAS] {10.1093/mnras/stu2674}, \href
  {http://adsabs.harvard.edu/abs/2015MNRAS.448...42L} {448, 42}

\bibitem[\protect\citeauthoryear{{Lamb} et~al.,}{{Lamb} et~al.}{2017}]{lamb17}
{Lamb} M.,  et~al., 2017, \mn@doi [MNRAS] {10.1093/mnras/stw2865}, \href
  {http://adsabs.harvard.edu/abs/2017MNRAS.465.3536L} {465, 3536}

\bibitem[\protect\citeauthoryear{{Lambert} \& {Ries}}{{Lambert} \&
  {Ries}}{1981}]{lambert81}
{Lambert} D.~L.,  {Ries} L.~M.,  1981, \mn@doi [ApJ] {10.1086/159147}, \href
  {http://adsabs.harvard.edu/abs/1981ApJ...248..228L} {248, 228}

\bibitem[\protect\citeauthoryear{{Larsson-Leander}}{{Larsson-Leander}}{1960}]{lar60}
{Larsson-Leander} G.,  1960, Stockholms Observatoriums Annaler, \href
  {http://adsabs.harvard.edu/abs/1960StoAn..20....9L} {20, 9}

\bibitem[\protect\citeauthoryear{{Lawler} \& {Dakin}}{{Lawler} \&
  {Dakin}}{1989}]{lawler89}
{Lawler} J.~E.,  {Dakin} J.~T.,  1989, \mn@doi [Journal of the Optical Society
  of America B Optical Physics] {10.1364/JOSAB.6.001457}, \href
  {http://adsabs.harvard.edu/abs/1989JOSAB...6.1457L} {6, 1457}

\bibitem[\protect\citeauthoryear{Lee}{Lee}{2015}]{lee15}
Lee J.-J.,  2015, plp: Version 2.0, \mn@doi{10.5281/zenodo.18579}, \url
  {https://doi.org/10.5281/zenodo.18579}

\bibitem[\protect\citeauthoryear{{Lindegren} et~al.,}{{Lindegren}
  et~al.}{2018}]{lindegren18}
{Lindegren} L.,  et~al., 2018, \mn@doi [A\&A] {10.1051/0004-6361/201832727},
  \href {http://adsabs.harvard.edu/abs/2018A%26A...616A...2L} {616, A2}

\bibitem[\protect\citeauthoryear{{Linden} et~al.,}{{Linden}
  et~al.}{2017}]{linden17}
{Linden} S.~T.,  et~al., 2017, \mn@doi [ApJ] {10.3847/1538-4357/aa6f17}, \href
  {http://adsabs.harvard.edu/abs/2017ApJ...842...49L} {842, 49}

\bibitem[\protect\citeauthoryear{{Lucatello}, {Masseron}, {Johnson},
  {Pignatari}  \& {Herwig}}{{Lucatello} et~al.}{2011}]{lucatello11}
{Lucatello} S.,  {Masseron} T.,  {Johnson} J.~A.,  {Pignatari} M.,   {Herwig}
  F.,  2011, \mn@doi [ApJ] {10.1088/0004-637X/729/1/40}, \href
  {http://adsabs.harvard.edu/abs/2011ApJ...729...40L} {729, 40}

\bibitem[\protect\citeauthoryear{{Mace} et~al.,}{{Mace} et~al.}{2016}]{mace16}
{Mace} G.,  et~al., 2016, in Ground-based and Airborne Instrumentation for
  Astronomy VI. p. 99080C, \mn@doi{10.1117/12.2232780}

\bibitem[\protect\citeauthoryear{{Maeder} \& {Meynet}}{{Maeder} \&
  {Meynet}}{1987}]{maeder87}
{Maeder} A.,  {Meynet} G.,  1987, A\&A, \href
  {http://adsabs.harvard.edu/abs/1987A%26A...182..243M} {182, 243}

\bibitem[\protect\citeauthoryear{{Magic}, {Weiss}  \& {Asplund}}{{Magic}
  et~al.}{2015}]{magic:15}
{Magic} Z.,  {Weiss} A.,   {Asplund} M.,  2015, \mn@doi [A\&A]
  {10.1051/0004-6361/201423760}, \href
  {http://adsabs.harvard.edu/abs/2015A%26A...573A..89M} {573, A89}

\bibitem[\protect\citeauthoryear{{Maiorca}, {Uitenbroek}, {Uttenthaler},
  {Randich}, {Busso}  \& {Magrini}}{{Maiorca} et~al.}{2014}]{maiorca14}
{Maiorca} E.,  {Uitenbroek} H.,  {Uttenthaler} S.,  {Randich} S.,  {Busso} M.,
   {Magrini} L.,  2014, \mn@doi [ApJ] {10.1088/0004-637X/788/2/149}, \href
  {http://adsabs.harvard.edu/abs/2014ApJ...788..149M} {788, 149}

\bibitem[\protect\citeauthoryear{{Majewski} et~al.,}{{Majewski}
  et~al.}{2017}]{majewski17}
{Majewski} S.~R.,  et~al., 2017, \mn@doi [AJ] {10.3847/1538-3881/aa784d}, \href
  {http://adsabs.harvard.edu/abs/2017AJ....154...94M} {154, 94}

\bibitem[\protect\citeauthoryear{{McLean} et~al.,}{{McLean}
  et~al.}{1998}]{mclean98}
{McLean} I.~S.,  et~al., 1998, in {Fowler} A.~M.,  ed.,  Proc. SPIE Vol. 3354,
  Infrared Astronomical Instrumentation. pp 566--578,
  \mn@doi{10.1117/12.317283}

\bibitem[\protect\citeauthoryear{{Mermilliod} \& {Mayor}}{{Mermilliod} \&
  {Mayor}}{1989}]{mermilliod89}
{Mermilliod} J.-C.,  {Mayor} M.,  1989, A\&A, \href
  {http://adsabs.harvard.edu/abs/1989A%26A...219..125M} {219, 125}

\bibitem[\protect\citeauthoryear{{Mishenina}, {Soubiran}, {Bienaym{\'e}},
  {Korotin}, {Belik}, {Usenko}  \& {Kovtyukh}}{{Mishenina}
  et~al.}{2008}]{mishenina08}
{Mishenina} T.~V.,  {Soubiran} C.,  {Bienaym{\'e}} O.,  {Korotin} S.~A.,
  {Belik} S.~I.,  {Usenko} I.~A.,   {Kovtyukh} V.~V.,  2008, \mn@doi [A\&A]
  {10.1051/0004-6361:200810360}, \href
  {http://adsabs.harvard.edu/abs/2008A%26A...489..923M} {489, 923}

\bibitem[\protect\citeauthoryear{{Moore}, {Minnaert}  \& {Houtgast}}{{Moore}
  et~al.}{1966}]{moore66}
{Moore} C.~E.,  {Minnaert} M.~G.~J.,   {Houtgast} J.,  1966, {The solar
  spectrum 2935 A to 8770 A}.
National Bureau of Standards Monograph, Washington: US Government Printing
  Office (USGPO)

\bibitem[\protect\citeauthoryear{{Mosumgaard}, {Ball}, {Silva Aguirre}, {Weiss}
   \& {Christensen-Dalsgaard}}{{Mosumgaard} et~al.}{2018}]{mosumgaard:18}
{Mosumgaard} J.~R.,  {Ball} W.~H.,  {Silva Aguirre} V.,  {Weiss} A.,
  {Christensen-Dalsgaard} J.,  2018, \mn@doi [MNRAS] {10.1093/mnras/sty1442},
  \href {http://adsabs.harvard.edu/abs/2018MNRAS.478.5650M} {478, 5650}

\bibitem[\protect\citeauthoryear{{Nataf} et~al.,}{{Nataf}
  et~al.}{2013}]{nataf13}
{Nataf} D.~M.,  et~al., 2013, \mn@doi [ApJ] {10.1088/0004-637X/769/2/88}, \href
  {http://adsabs.harvard.edu/abs/2013ApJ...769...88N} {769, 88}

\bibitem[\protect\citeauthoryear{{Oliva} et~al.,}{{Oliva}
  et~al.}{2012}]{oliva12}
{Oliva} E.,  et~al., 2012, in Ground-based and Airborne Instrumentation for
  Astronomy IV. p. 84463T, \mn@doi{10.1117/12.925274}

\bibitem[\protect\citeauthoryear{{Origlia} \& {Rich}}{{Origlia} \&
  {Rich}}{2004}]{origlia04}
{Origlia} L.,  {Rich} R.~M.,  2004, \mn@doi [AJ] {10.1086/420704}, \href
  {http://adsabs.harvard.edu/abs/2004AJ....127.3422O} {127, 3422}

\bibitem[\protect\citeauthoryear{{Origlia}, {Rich}  \& {Castro}}{{Origlia}
  et~al.}{2002}]{origlia02}
{Origlia} L.,  {Rich} R.~M.,   {Castro} S.,  2002, \mn@doi [AJ]
  {10.1086/338897}, \href {http://adsabs.harvard.edu/abs/2002AJ....123.1559O}
  {123, 1559}

\bibitem[\protect\citeauthoryear{{Origlia}, {Valenti}  \& {Rich}}{{Origlia}
  et~al.}{2005}]{origlia05}
{Origlia} L.,  {Valenti} E.,   {Rich} R.~M.,  2005, \mn@doi [MNRAS]
  {10.1111/j.1365-2966.2004.08529.x}, \href
  {http://adsabs.harvard.edu/abs/2005MNRAS.356.1276O} {356, 1276}

\bibitem[\protect\citeauthoryear{{Origlia}, {Valenti}, {Rich}  \&
  {Ferraro}}{{Origlia} et~al.}{2006}]{origlia06}
{Origlia} L.,  {Valenti} E.,  {Rich} R.~M.,   {Ferraro} F.~R.,  2006, \mn@doi
  [ApJ] {10.1086/504829}, \href
  {http://adsabs.harvard.edu/abs/2006ApJ...646..499O} {646, 499}

\bibitem[\protect\citeauthoryear{{Origlia}, {Valenti}  \& {Rich}}{{Origlia}
  et~al.}{2008}]{origlia08}
{Origlia} L.,  {Valenti} E.,   {Rich} R.~M.,  2008, \mn@doi [MNRAS]
  {10.1111/j.1365-2966.2008.13492.x}, \href
  {http://adsabs.harvard.edu/abs/2008MNRAS.388.1419O} {388, 1419}

\bibitem[\protect\citeauthoryear{{Origlia} et~al.,}{{Origlia}
  et~al.}{2013}]{origlia13}
{Origlia} L.,  et~al., 2013, \mn@doi [A\&A] {10.1051/0004-6361/201322586},
  \href {http://adsabs.harvard.edu/abs/2013A%26A...560A..46O} {560, A46}

\bibitem[\protect\citeauthoryear{{Origlia} et~al.,}{{Origlia}
  et~al.}{2016}]{origlia16}
{Origlia} L.,  et~al., 2016, \mn@doi [A\&A] {10.1051/0004-6361/201526649},
  \href {http://adsabs.harvard.edu/abs/2016A%26A...585A..14O} {585, A14}

\bibitem[\protect\citeauthoryear{{Pancino}}{{Pancino}}{2018}]{pancino18}
{Pancino} E.,  2018, \mn@doi [\aap] {10.1051/0004-6361/201732351}, \href
  {http://adsabs.harvard.edu/abs/2018A%26A...614A..80P} {614, A80}

\bibitem[\protect\citeauthoryear{{Park} et~al.,}{{Park} et~al.}{2014}]{park14}
{Park} C.,  et~al., 2014, in Society of Photo-Optical Instrumentation Engineers
  (SPIE) Conference Series. p.~1, \mn@doi{10.1117/12.2056431}

\bibitem[\protect\citeauthoryear{{Paxton}, {Bildsten}, {Dotter}, {Herwig},
  {Lesaffre}  \& {Timmes}}{{Paxton} et~al.}{2011}]{paxton:11}
{Paxton} B.,  {Bildsten} L.,  {Dotter} A.,  {Herwig} F.,  {Lesaffre} P.,
  {Timmes} F.,  2011, \mn@doi [ApJS] {10.1088/0067-0049/192/1/3}, \href
  {http://adsabs.harvard.edu/abs/2011ApJS..192....3P} {192, 3}

\bibitem[\protect\citeauthoryear{{Paxton} et~al.,}{{Paxton}
  et~al.}{2013}]{paxton:13}
{Paxton} B.,  et~al., 2013, \mn@doi [ApJS] {10.1088/0067-0049/208/1/4}, \href
  {http://adsabs.harvard.edu/abs/2013ApJS..208....4P} {208, 4}

\bibitem[\protect\citeauthoryear{{Pilachowski} \& {Pace}}{{Pilachowski} \&
  {Pace}}{2015}]{pilachowski15}
{Pilachowski} C.~A.,  {Pace} C.,  2015, \mn@doi [AJ]
  {10.1088/0004-6256/150/3/66}, \href
  {http://adsabs.harvard.edu/abs/2015AJ....150...66P} {150, 66}

\bibitem[\protect\citeauthoryear{{Randich}, {Sestito}, {Primas}, {Pallavicini}
  \& {Pasquini}}{{Randich} et~al.}{2006}]{randich06}
{Randich} S.,  {Sestito} P.,  {Primas} F.,  {Pallavicini} R.,   {Pasquini} L.,
  2006, \mn@doi [A\&A] {10.1051/0004-6361:20054291}, \href
  {http://adsabs.harvard.edu/abs/2006A%26A...450..557R} {450, 557}

\bibitem[\protect\citeauthoryear{{Rothman} et~al.,}{{Rothman}
  et~al.}{2013}]{rothman13}
{Rothman} L.~S.,  et~al., 2013, \mn@doi [JQSRTJ. Quant. Spectrosc. Radiat.
  Transfer] {10.1016/j.jqsrt.2013.07.002}, \href
  {http://adsabs.harvard.edu/abs/2013JQSRT.130....4R} {130, 4}

\bibitem[\protect\citeauthoryear{{Roxburgh}}{{Roxburgh}}{1989}]{roxburgh:89}
{Roxburgh} I.~W.,  1989, A\&A, \href
  {http://adsabs.harvard.edu/abs/1989A%26A...211..361R} {211, 361}

\bibitem[\protect\citeauthoryear{{Ruffoni}, {Den Hartog}, {Lawler}, {Brewer},
  {Lind}, {Nave}  \& {Pickering}}{{Ruffoni} et~al.}{2014}]{ruffoni14}
{Ruffoni} M.~P.,  {Den Hartog} E.~A.,  {Lawler} J.~E.,  {Brewer} N.~R.,  {Lind}
  K.,  {Nave} G.,   {Pickering} J.~C.,  2014, \mn@doi [MNRAS]
  {10.1093/mnras/stu780}, \href
  {http://adsabs.harvard.edu/abs/2014MNRAS.441.3127R} {441, 3127}

\bibitem[\protect\citeauthoryear{{Salaris} \& {Cassisi}}{{Salaris} \&
  {Cassisi}}{2015}]{salaris:15}
{Salaris} M.,  {Cassisi} S.,  2015, \mn@doi [A\&A]
  {10.1051/0004-6361/201525812}, \href
  {http://adsabs.harvard.edu/abs/2015A%26A...577A..60S} {577, A60}

\bibitem[\protect\citeauthoryear{{Schaller}, {Schaerer}, {Meynet}  \&
  {Maeder}}{{Schaller} et~al.}{1992}]{schaller:92}
{Schaller} G.,  {Schaerer} D.,  {Meynet} G.,   {Maeder} A.,  1992, A\&ASS,
  \href {http://adsabs.harvard.edu/abs/1992A%26AS...96..269S} {96, 269}

\bibitem[\protect\citeauthoryear{{Shetrone}}{{Shetrone}}{2003}]{shetrone03}
{Shetrone} M.~D.,  2003, \mn@doi [ApJL] {10.1086/374262}, \href
  {http://adsabs.harvard.edu/abs/2003ApJ...585L..45S} {585, L45}

\bibitem[\protect\citeauthoryear{{Simmerer}, {Sneden}, {Cowan}, {Collier},
  {Woolf}  \& {Lawler}}{{Simmerer} et~al.}{2004}]{simmerer04}
{Simmerer} J.,  {Sneden} C.,  {Cowan} J.~J.,  {Collier} J.,  {Woolf} V.~M.,
  {Lawler} J.~E.,  2004, \mn@doi [ApJ] {10.1086/424504}, \href
  {http://adsabs.harvard.edu/abs/2004ApJ...617.1091S} {617, 1091}

\bibitem[\protect\citeauthoryear{{Smiljanic}, {Donati}, {Bragaglia}, {Lemasle}
  \& {Romano}}{{Smiljanic} et~al.}{2018a}]{smiljanic18}
{Smiljanic} R.,  {Donati} P.,  {Bragaglia} A.,  {Lemasle} B.,   {Romano} D.,
  2018a, \mn@doi [\aap] {10.1051/0004-6361/201832877}, \href
  {http://adsabs.harvard.edu/abs/2018A%26A...616A.112S} {616, A112}

\bibitem[\protect\citeauthoryear{{Smiljanic} et~al.,}{{Smiljanic}
  et~al.}{2018b}]{smiljanic:18}
{Smiljanic} R.,  et~al., 2018b, \mn@doi [A\&A] {10.1051/0004-6361/201833027},
  \href {http://adsabs.harvard.edu/abs/2018A%26A...617A...4S} {617, A4}

\bibitem[\protect\citeauthoryear{{Sneden}}{{Sneden}}{1973}]{sneden73}
{Sneden} C.,  1973, \mn@doi [ApJ] {10.1086/152374}, \href
  {http://adsabs.harvard.edu/abs/1973ApJ...184..839S} {184, 839}

\bibitem[\protect\citeauthoryear{{Sneden}, {Cowan}  \& {Gallino}}{{Sneden}
  et~al.}{2008}]{sneden08}
{Sneden} C.,  {Cowan} J.~J.,   {Gallino} R.,  2008, \mn@doi [ARA\&A]
  {10.1146/annurev.astro.46.060407.145207}, \href
  {http://adsabs.harvard.edu/abs/2008ARA%26A..46..241S} {46, 241}

\bibitem[\protect\citeauthoryear{{Souto} et~al.,}{{Souto}
  et~al.}{2016}]{souto16}
{Souto} D.,  et~al., 2016, \mn@doi [ApJ] {10.3847/0004-637X/830/1/35}, \href
  {http://adsabs.harvard.edu/abs/2016ApJ...830...35S} {830, 35}

\bibitem[\protect\citeauthoryear{{Souto} et~al.,}{{Souto}
  et~al.}{2018}]{souto18}
{Souto} D.,  et~al., 2018, \mn@doi [ApJ] {10.3847/1538-4357/aab612}, \href
  {http://adsabs.harvard.edu/abs/2018ApJ...857...14S} {857, 14}

\bibitem[\protect\citeauthoryear{{Stancliffe}, {Fossati}, {Passy}  \&
  {Schneider}}{{Stancliffe} et~al.}{2015}]{stancliffe15}
{Stancliffe} R.~J.,  {Fossati} L.,  {Passy} J.-C.,   {Schneider} F.~R.~N.,
  2015, \mn@doi [A\&A] {10.1051/0004-6361/201425126}, \href
  {http://adsabs.harvard.edu/abs/2015A%26A...575A.117S} {575, A117}

\bibitem[\protect\citeauthoryear{{Stanek}, {Udalski}, {Szyma{\'N}ski},
  {Ka{\L}u{\.Z}ny}, {Kubiak}, {Mateo}  \& {Krzemi{\'N}ski}}{{Stanek}
  et~al.}{1997}]{stanek97}
{Stanek} K.~Z.,  {Udalski} A.,  {Szyma{\'N}ski} M.,  {Ka{\L}u{\.Z}ny} J.,
  {Kubiak} Z.~M.,  {Mateo} M.,   {Krzemi{\'N}ski} W.,  1997, \mn@doi [ApJ]
  {10.1086/303702}, \href {http://adsabs.harvard.edu/abs/1997ApJ...477..163S}
  {477, 163}

\bibitem[\protect\citeauthoryear{{Stetson}}{{Stetson}}{2000}]{stetson00}
{Stetson} P.~B.,  2000, \mn@doi [PASP] {10.1086/316595}, \href
  {http://adsabs.harvard.edu/abs/2000PASP..112..925S} {112, 925}

\bibitem[\protect\citeauthoryear{{Takeda}, {Zhao}, {Chen}, {Qiu}  \&
  {Takada-Hidai}}{{Takeda} et~al.}{2002}]{takeda02}
{Takeda} Y.,  {Zhao} G.,  {Chen} Y.-Q.,  {Qiu} H.-M.,   {Takada-Hidai} M.,
  2002, \mn@doi [PASJ] {10.1093/pasj/54.2.275}, \href
  {http://adsabs.harvard.edu/abs/2002PASJ...54..275T} {54, 275}

\bibitem[\protect\citeauthoryear{{Takeda}, {Kaneko}, {Matsumoto}, {Oshino},
  {Ito}  \& {Shibuya}}{{Takeda} et~al.}{2009}]{takeda09}
{Takeda} Y.,  {Kaneko} H.,  {Matsumoto} N.,  {Oshino} S.,  {Ito} H.,
  {Shibuya} T.,  2009, \mn@doi [PASJ] {10.1093/pasj/61.3.563}, \href
  {http://adsabs.harvard.edu/abs/2009PASJ...61..563T} {61, 563}

\bibitem[\protect\citeauthoryear{{Thygesen} et~al.,}{{Thygesen}
  et~al.}{2014}]{thygesan14}
{Thygesen} A.~O.,  et~al., 2014, \mn@doi [A\&A] {10.1051/0004-6361/201424533},
  \href {http://adsabs.harvard.edu/abs/2014A%26A...572A.108T} {572, A108}

\bibitem[\protect\citeauthoryear{{Tull}}{{Tull}}{1998}]{tull98}
{Tull} R.~G.,  1998, in {D'Odorico} S.,  ed.,  Proc. SPIE Vol. 3355, Optical
  Astronomical Instrumentation. pp 387--398, \mn@doi{10.1117/12.316774}

\bibitem[\protect\citeauthoryear{{Valenti}, {Origlia}  \& {Rich}}{{Valenti}
  et~al.}{2011}]{valenti11}
{Valenti} E.,  {Origlia} L.,   {Rich} R.~M.,  2011, \mn@doi [MNRAS]
  {10.1111/j.1365-2966.2011.18580.x}, \href
  {http://adsabs.harvard.edu/abs/2011MNRAS.414.2690V} {414, 2690}

\bibitem[\protect\citeauthoryear{{Valenti}, {Origlia}, {Mucciarelli}  \&
  {Rich}}{{Valenti} et~al.}{2015}]{valenti15}
{Valenti} E.,  {Origlia} L.,  {Mucciarelli} A.,   {Rich} R.~M.,  2015, \mn@doi
  [A\&A] {10.1051/0004-6361/201424888}, \href
  {http://adsabs.harvard.edu/abs/2015A%26A...574A..80V} {574, A80}

\bibitem[\protect\citeauthoryear{{Valle}, {Dell'Omodarme}, {Prada Moroni}  \&
  {Degl'Innocenti}}{{Valle} et~al.}{2017}]{valle:17}
{Valle} G.,  {Dell'Omodarme} M.,  {Prada Moroni} P.~G.,   {Degl'Innocenti} S.,
  2017, \mn@doi [A\&A] {10.1051/0004-6361/201628240}, \href
  {http://adsabs.harvard.edu/abs/2017A%26A...600A..41V} {600, A41}

\bibitem[\protect\citeauthoryear{{VandenBerg}, {Bergbusch}  \&
  {Dowler}}{{VandenBerg} et~al.}{2006}]{vandenberg:06}
{VandenBerg} D.~A.,  {Bergbusch} P.~A.,   {Dowler} P.~D.,  2006, \mn@doi [ApJS]
  {10.1086/498451}, \href {http://adsabs.harvard.edu/abs/2006ApJS..162..375V}
  {162, 375}

\bibitem[\protect\citeauthoryear{{VandenBerg}, {Edvardsson}, {Eriksson}  \&
  {Gustafsson}}{{VandenBerg} et~al.}{2008}]{vandenberg:08}
{VandenBerg} D.~A.,  {Edvardsson} B.,  {Eriksson} K.,   {Gustafsson} B.,  2008,
  \mn@doi [ApJ] {10.1086/521600}, \href
  {http://adsabs.harvard.edu/abs/2008ApJ...675..746V} {675, 746}

\bibitem[\protect\citeauthoryear{{VandenBerg}, {Brogaard}, {Leaman}  \&
  {Casagrande}}{{VandenBerg} et~al.}{2013}]{vandenberg:13}
{VandenBerg} D.~A.,  {Brogaard} K.,  {Leaman} R.,   {Casagrande} L.,  2013,
  \mn@doi [ApJ] {10.1088/0004-637X/775/2/134}, \href
  {http://adsabs.harvard.edu/abs/2013ApJ...775..134V} {775, 134}

\bibitem[\protect\citeauthoryear{{VandenBerg}, {Bergbusch}, {Ferguson}  \&
  {Edvardsson}}{{VandenBerg} et~al.}{2014}]{vandenberg:14}
{VandenBerg} D.~A.,  {Bergbusch} P.~A.,  {Ferguson} J.~W.,   {Edvardsson} B.,
  2014, \mn@doi [ApJ] {10.1088/0004-637X/794/1/72}, \href
  {http://adsabs.harvard.edu/abs/2014ApJ...794...72V} {794, 72}

\bibitem[\protect\citeauthoryear{{Walker}}{{Walker}}{1958}]{walker58}
{Walker} M.~F.,  1958, \mn@doi [ApJ] {10.1086/146570}, \href
  {http://adsabs.harvard.edu/abs/1958ApJ...128..562W} {128, 562}

\bibitem[\protect\citeauthoryear{{Wallerstein} \& {Helfer}}{{Wallerstein} \&
  {Helfer}}{1959}]{wallerstein59}
{Wallerstein} G.,  {Helfer} H.~L.,  1959, \mn@doi [ApJ] {10.1086/146669}, \href
  {http://adsabs.harvard.edu/abs/1959ApJ...129..720W} {129, 720}

\bibitem[\protect\citeauthoryear{{Yuk} et~al.,}{{Yuk} et~al.}{2010}]{yuk10}
{Yuk} I.-S.,  et~al., 2010, in Society of Photo-Optical Instrumentation
  Engineers (SPIE) Conference Series. p.~1, \mn@doi{10.1117/12.856864}

\bibitem[\protect\citeauthoryear{{de Laverny} \& {Recio-Blanco}}{{de Laverny}
  \& {Recio-Blanco}}{2013}]{delaverny13}
{de Laverny} P.,  {Recio-Blanco} A.,  2013, \mn@doi [A\&A]
  {10.1051/0004-6361/201322222}, \href
  {http://adsabs.harvard.edu/abs/2013A%26A...560A..74D} {560, A74}

\makeatother
\end{thebibliography}


\bsp	
\label{lastpage}
\end{document}